<u>**Title:**</u> **Decades matter: Agricultural diversification increases financial profitability, biodiversity, and ecosystem services over time.**

Authors: Estelle Raveloaritiana[1,2,*,] Thomas Cherico Wanger [1,2,3,4,5*]

[1] Sustainable Agricultural Systems & Engineering Laboratory, School of Engineering, Westlake University, Hangzhou, China.
[2] Key Laboratory of Coastal Environment and Resources of Zhejiang Province, Westlake University, Hangzhou, China.
[3] Faculty of Science, Technology & Medicine, University of Luxembourg, Luxembourg
[4] Global Agroforestry Network, Hangzhou, China
[5] China Rice Network, Hangzhou, China

*Corresponding authors: eraveloaritiana@gmail.com (ER); tomcwanger@gmail.com (TCW)

**Abstract (194/200 words maximum)**

Sustainable agriculture in the 21st century requires the production of sufficient food while reducing the environmental impact and safeguarding human livelihoods [1,2]. Many studies have confirmed agricultural diversification practices such as intercropping, organic farming and soil inoculations as a suitable pathway to achieve these goals [3,4] but long-term viability of socioeconomic and ecological benefits is uncertain [5,6]. Here, we quantified the long-term effects of agricultural diversification practices on socioeconomic and ecological benefits based on 50 years of data from 184 meta-analyses and 4,260 effect sizes. We showed that, with neutral crop yield over time, financial profitability, most variables related to biological communities, all aspects of soil quality, and carbon sequestration benefits increased by up to 2823% over 20 years of practice. Non-crop diversification practices and the use of organic amendments increased benefits by up to 2000% after 50 years. A trade-off analysis between yield and other services showed win-win outcomes during the first 25 years. Our synthesis provides the urgently needed evidence for farmers and other decision-makers that diversification increases long-term profitability, biodiversity, and climate mitigation benefits, and therefore, allows upscaling diversification for climate change mitigation and global food system transformation [7].

<u>**Keywords:**</u> diversified farming, ecosystem services, biodiversity, yield, financial profitability, trade-off over time, practices duration

*Main*

Transforming the global food system requires nature-positive production systems [4,8] based on agricultural diversification to achieve sustainable food production [9–11]. Agricultural diversification is based on farming practices, such as crop rotation, agroforestry, or organic farming, that purposely integrate functional diversity into production systems to enhance or maintain the services essential to crop production [3]. Several syntheses have confirmed the socioeconomic (i.e., yield and profitability) and environmental (i.e., biological communities, soil quality, and climate mitigation) benefits of agricultural diversification practices for crop production worldwide [5,6]. In Asia, for example, rice crop diversification practices reduce pesticide use and increase biocontrol and crop yields [12]. These benefits have incentivized stakeholder groups to implement agricultural diversification across different countries. In China, agricultural diversification was recently promoted into the country's major agricultural policies [13]. In Europe and the USA, several agricultural diversification practices, such as organic farming and crop/non-crop diversification, are advocated in agricultural policies for biodiversity conservation and climate change mitigation [14–16]. Despite these successes, the long-term effects of agricultural diversification on socioeconomic and environmental factors remain unquantified and impede large-scale implementation by risk-averse farmers and promotion through policy makers.

Stakeholder groups require long-term evidence that diversified agricultural production can maintain and ideally improve livelihoods, economy, and the environment to mitigate climate change and external stressors[17–19]. For instance, in staple foods like rice, the socioeconomic and environmental benefits of diversified rice farming are now well quantified [20], but implementation success will depend on understanding temporal dynamics of such benefits [17]. In cash crops like cocoa, smallholder farmers have made long-term investments into their tree plantations and may be reluctant to transition towards diversified agroforestry systems without understanding long-term effects on profitability and yields [21]. More generally, global food insecurity requires policy decisions that assure reaching the Sustainable Development Goals now more than ever [7,22,23]. Moving beyond non-temporal analyses and quantifying the long-term effects of agricultural diversification practices on socioeconomic and environmental factors - and the trade-offs between them is, hence, decisive to the successful implementation of a global food system transformation [22,24,25].

Here, we conducted a global synthesis of meta-analyses to quantify the effects of agricultural diversification practices on socioeconomic and environmental factors for up to 50 years. Based on data from 184 meta-analyses with 4,260 effect sizes based on 17,989 original studies (Supplementary Figure 1), we quantified the long-term effects on socioeconomic factors (crop yield and financial profitability), biological communities (biodiversity, pollination, and pest control), soil quality (water regulation, soil fertility, and nutrient cycling), and climate change mitigation (carbon sequestration and climate regulation). We also analysed the temporal trade-offs between crop yield and ecosystem services (Methods).

### Socioeconomic, biological community, soil quality, and climate change mitigation benefits mostly increase over time

Our second-order meta-regression models showed that agricultural diversification practices significantly increased benefits for seven out of ten socio-economic and environmental factors in the long term (Fig. 1, Extended Data Table 1). The analysis of socioeconomic factors showed that while agricultural diversification did not reduce crop yield over time (Fig. 1), financial profitability increased significantly by 172% over 20 years compared to short-term effects (< 3 years, Extended Data Fig. 2, Extended Data Table 2). These findings show that the crop yield of diversified farming systems stabilizes over time [26,27], while farmers' income improves and, hence, highlighting the socio-economic viability of agricultural diversification in the long run [9].

For biological communities, agricultural diversification significantly increased biodiversity and pollination benefits by 231% and 2823%, respectively, over 20 years (Fig. 1, Extended Data Table 2). These patterns provide evidence that agricultural diversification can play an important role not only in bending the curve of biodiversity losses from agriculture [28] but also in improving pollination services [29], which is highly important for global food security [30]. However, the effects of diversification on pest control decreased over time (Fig. 1, Extended Data Table 1-2), but with an increased probability of neutral effects over the years (Extended Data Fig. 1). Diversified farming may, hence, provide concomitant benefits to pest control agents and pests [31], which leads to overall negative trends.

All effect sizes of soil quality and climate change mitigation, except for climate regulation, were significantly positive but did not significantly change over time (Fig.1). These patterns suggest that benefits of agricultural diversification for water regulation, soil fertility, nutrient cycling, and carbon sequestration are maintained in the long run. The transformation from simple monocultures to diversified farming systems, therefore, can improve soil quality and carbon stocks in agricultural lands sustainably [32,33].

Overall, our results provide evidence that agricultural diversification can help to achieve sustainable and climate resilient food production [4,34] with financial profitability, biodiversity and pollination being 100%, 56% and 1164%, respectively, higher after 20 years compared to non-temporal analysis results (Extended Figure 2, Extended Table 2). Our analyses show that non-temporal analyses vastly underestimate the full potential that agricultural diversification can bring for socioeconomic and environmental factors after just two decades [22,34].

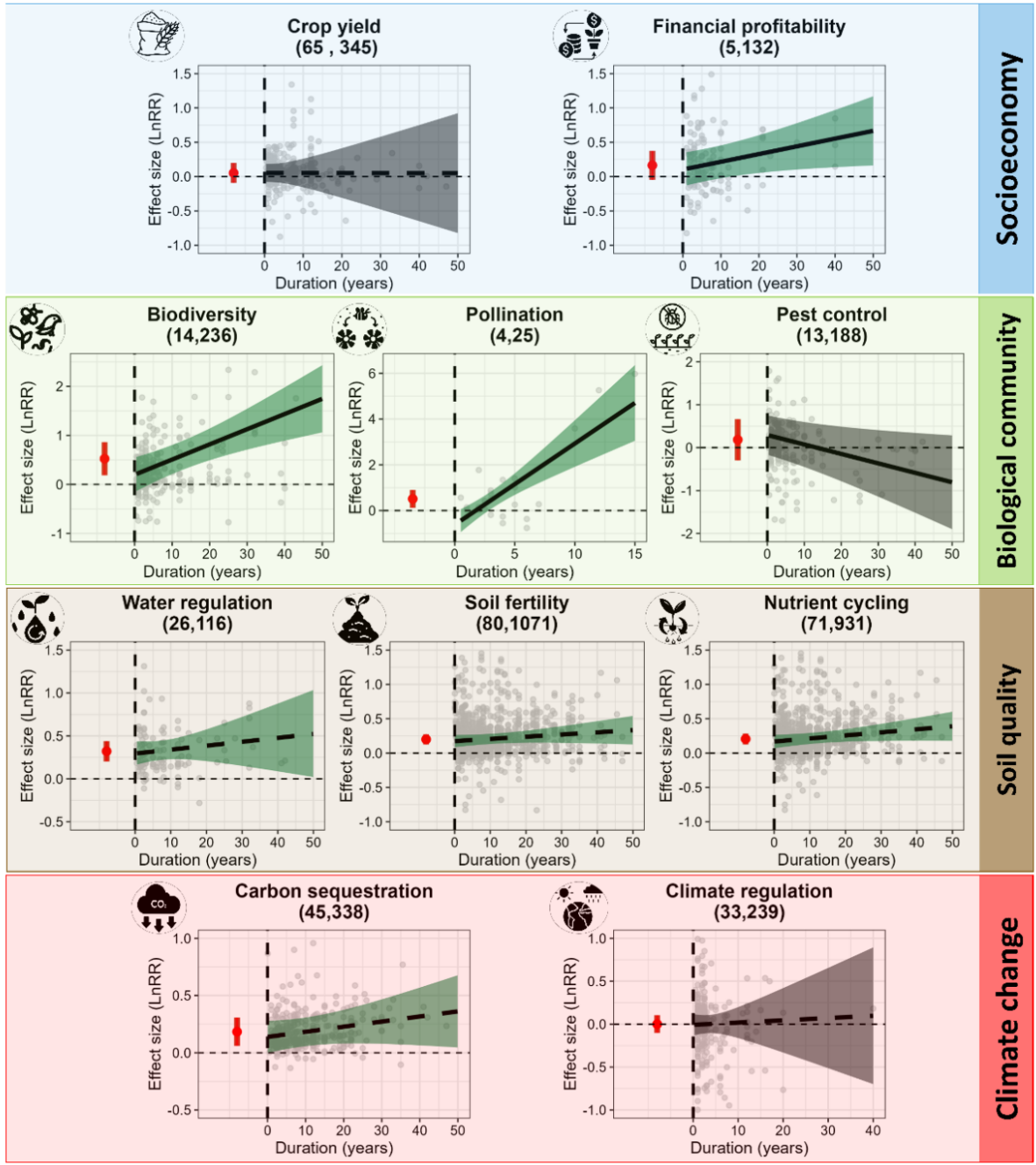

Fig. 1| Effects of agricultural diversification on socioeconomic, biological community, soil quality, and climate change-related variables over time compared to non-temporal effects. Red dot and bars represent non-temporal means with 95% confidence intervals (CI), respectively, *sensu* Tamburini et al. 2020. Respective solid or dashed lines inside ribbons represent significant and non-significant change of effect sizes (LnRR) over time; ribbons are the 95% CI. Ribbon colours: grey = effects of diversification practices are neutral; green = significant positive effect of diversification. In brackets are the number of studies and effect sizes.

Our meta-regression analyses on the effects of six diversification practices on socioeconomic factors, biological community, soil quality, and climate change categories showed that almost half of the response variables became significantly positive or increased over time (dark green in Fig. 2, Extended Data Table 3-4). Non-crop diversification and organic amendments increased response variables in all four categories over time (Fig. 2, Extended Data Table 4), which provides realistic opportunities to address the twin challenges of biodiversity loss [28,35] and climate mitigation [36,37]. In addition, the long-term maintenance or significant improvement of soil fertility and nutrient cycling shows enhancement of soil quality [36,38,39]. Long-term benefits of organic amendments highlight the potential of this practice to reduce chemical inputs, and hence, pollution and health repercussions [40,41]. Therefore, these practices constitute the best options for shifting away from unsustainable and towards more biodiversity and nature-friendly farming practices [9,22].

Among all the types of agricultural diversification practices, crop diversification was the only practice that significantly increased financial profitability and resulted in neutral yields over time (Fig. 2, Extended Data Table 4). Crop diversification can, hence, improve farmers' income over time [42] and ensure the long-term stability of crop yield [27]. While pest control on crop diversification decreased significantly with time, there was no significant difference compared to conventional agriculture for up to 50 years (Fig. 1, Extended Data Table 4). This result indicates that some known benefits for instance of crop rotation on pest control may be counteracted [43], where environmental complexity increases pest incidence and herbivory richness [43–45]. Reduced tillage showed slight yield increases over time but significant long-term benefits for carbon sequestration, while organic farming did not provide significant long-term benefits (Fig. 2, Extended Data Table 4). Overall, mainstreaming agricultural diversification and the combination of various practices may be a very effective and long-term strategy for a global food systems transformation [11]. For instance, non-crop diversification strategies seem to hold most benefits as an individual practice, while organic farming shows an incremental increase in biodiversity benefits over time, but has consistently lower yields (but see [46]). An effective combination of practices may help to transform some of the neutral into positive long-term benefits [6].

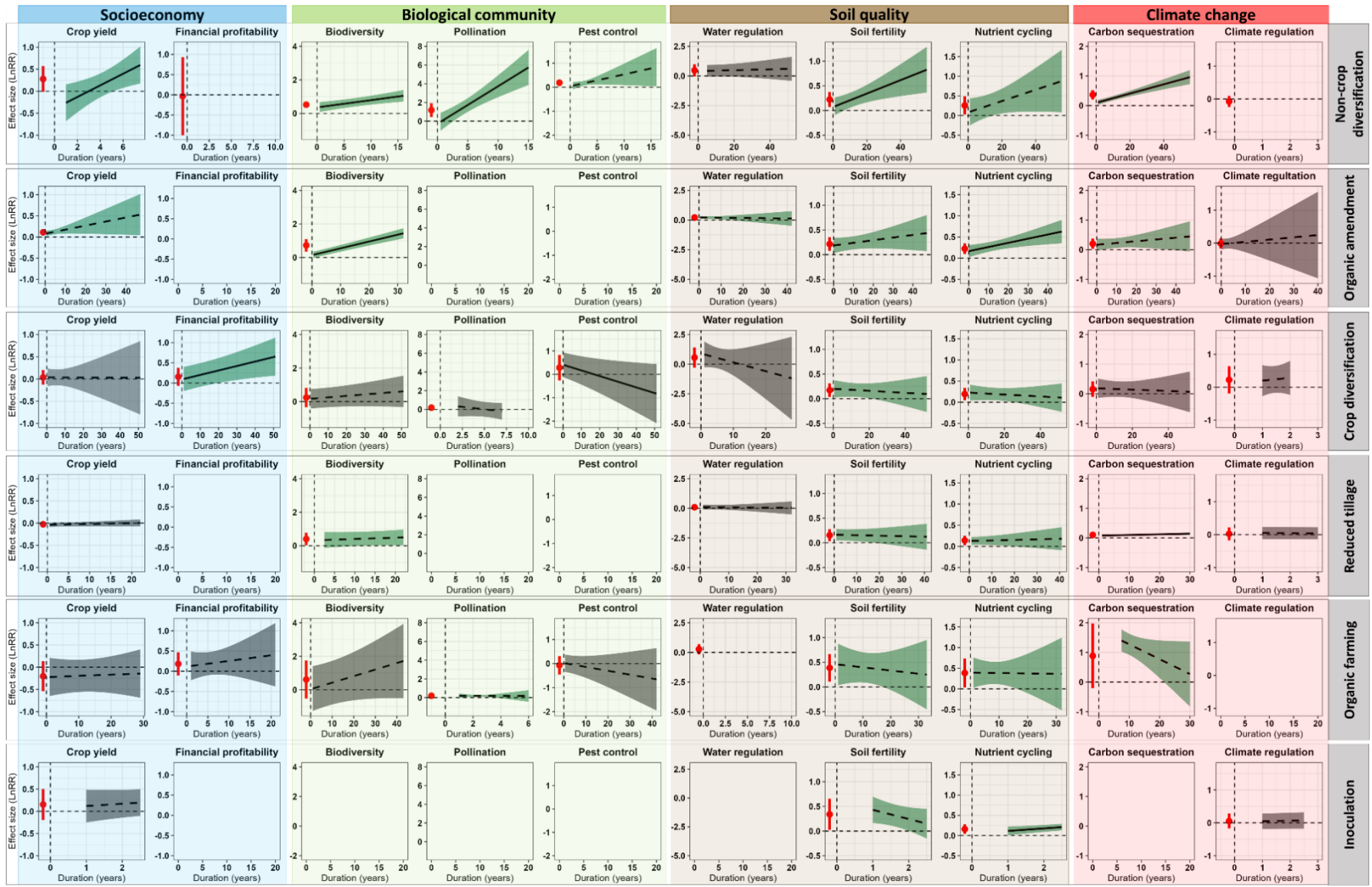

Fig. 2| Effects of individual diversification practices on socioeconomic, biological community, soil quality, and climate change-related variables over time compared to non-temporal effects. Red dot and bars represent means from non-temporal analyses with 95% confidence intervals (CI), respectively, *sensu* Tamburini et al. 2020. Respective solid or dashed lines inside ribbons represent significant and non-significant change of effect sizes (LnRR) over time; ribbons are the 95% CI. Ribbon colours: grey = effects are neutral over time; green = significant positive effect of diversification. For non-crop diversification, financial profitability and climate regulation predictions cover three years and one year, respectively. For organic farming, the water regulation predictions cover 33 years. Empty panels indicate research gaps due to data unavailability.

*Long term trade-offs between crop yield and other ecosystem services*

We predicted trade-offs between crop yield and other services over time based on hierarchical regression models, in which all effect sizes were divided into loss and win outcomes (Methods). Overall, we detected a win-win situation between crop yield and all other services, with a probability higher than 60% within the first 25 years of agricultural diversification practices, and then a lose yield-win ecosystem services was the most likely outcome (Fig. 3A, Extended Data Table 5).

When examining the trade-off between yield and individual services, we found that increasing win-win situations with biodiversity, water regulation, and climate regulation were predominant (Fig. 3B-J, Extended Data Table 5). These results show that the positive and desirable outcomes of agricultural diversification practices not only remain but also become more probable over time [8]. Our analyses also showed win-win situations with financial profitability, soil fertility, nutrient cycling, and carbon sequestration for 5, 24, 25 and 15 years, respectively (Fig. 3) and lose yield-win ecosystem service situations becoming more common afterwards. Our results indicate that compared to monocultures and conventional agriculture, agricultural diversification can improve crop yield while increasing farmers' income, soil nutrients, and capturing and storing carbon from the atmosphere, thus achieving nature-positive systems and equitable livelihoods [4,47,48]. Government agri-environmental schemes should, hence, compensate for lower yields in diversified systems [49].

We also found a predominance of lose yield-win pest control and pollination relationships over the first ten years (Fig. 3), indicating that increases in pollination and pest control services do not always translate into higher crop yields. It is noteworthy, however, that while there is not necessarily a yield increase with pollination, 70% of all crops globally depend on natural pollination; hence, pollination is critical for yield stability [30,50]. In addition, some studies have shown that pollination and pest control services can increase yield quality, thus improving the market value of crop yield [51]. Our analysis also revealed a temporal increase of the lose-win relationship between crop yield and carbon sequestration. Thus, climate financing for diversified agriculture will account for such carbon-yield trade-offs and requires even more advocacy for diversification and agroecology at climate summit discussions.

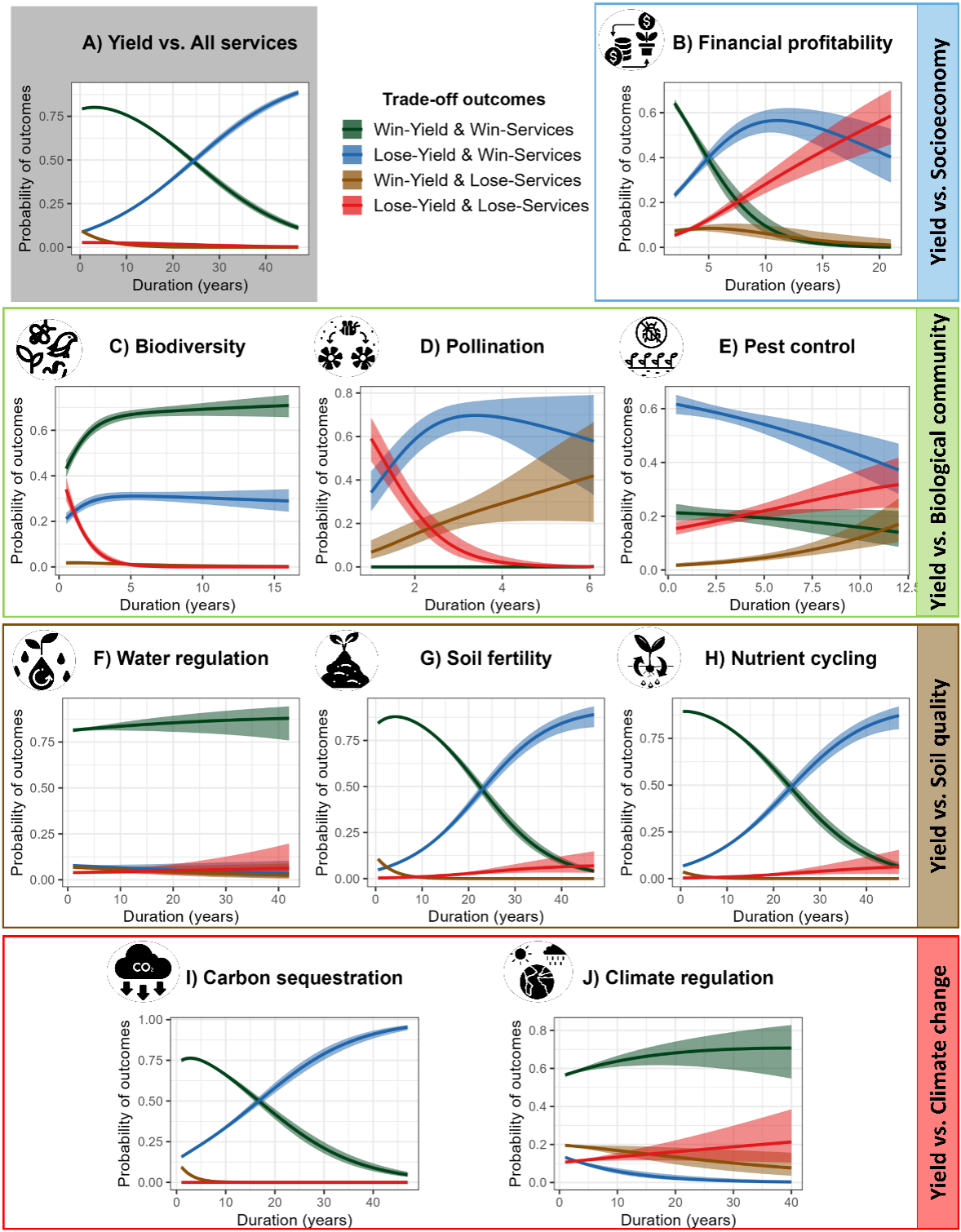

Fig. 3| Trade-offs for agricultural diversification implementation over time: crop yield vs.

overall (A) and individual ecosystem services in the four categories of socioeconomy (B), biological community (C-E), soil quality (F-H), and climate change (I-J).

***Research gaps***

Our work quantifies diversification effects for up to 50 years and allows us to identify temporal research gaps for the first time. First, short-term meta-analyses (0-10 years) on diversification effects on socioeconomic, biological community, soil quality, and climate change related variables were almost twice more common (137studies and 2479 effect sizes) than long-term (>10years) meta-analyses (74 studies and 1140 effect sizes). Likely, primary studies with long-term data are lacking, as funding opportunities for studies or experiments of >10 years are limited [52], but critically important to understand, for instance, diversification effects on pollination, climate regulation, and financial profitability (see Fig. 1 and Extended Data Fig. 3). Second, our trade-off analysis contained less data (43 studies, 1448 effect sizes) than the general meta-regressions (148 studies, 3619 effect sizes, Extended data Fig. 3-4). More studies are needed that concomitantly focus on diversification effects on yield or other services. Lastly, for individual diversification practices, we found no data on the effects of organic amendments and reduced tillage on financial profitability, pollination, and pest control over time (Fig. 2). Nevertheless, the use of these two practices is highly prevalent and advocated globally [34,38]. There were also no data on organic farming effects on climate regulation and almost no effect sizes for long-term inoculation effects (see Fig. 2). Overall, syntheses based on long-term studies require innovative funding mechanisms from various sources for ultimately managing all trade-offs in diversified farming systems to maintain food security, alleviate poverty, and reduce environmental impacts [53,54].

***A long-term perspective to implement agricultural diversification.***

We show that non-temporal analyses have underestimated the benefits of agricultural diversification by up to 1164%, particularly for financial profitability, biodiversity, and pollination. Our work is the first quantification of the long-term benefits of agricultural diversification that can inform researchers, policymakers, and practitioners on urgently needed implementation strategies[8].

For the research community, it is important to focus on long-term projects that can disentangle the complex relationships between yield and various ecosystem services linked to biodiversity loss and climate change [9,17]. Intuitively, a long-term focus will also address many remaining short-term research gaps on the way [5]. Clearly, addressing trade-offs between different ecosystem services requires multidisciplinary collaborations and international partnerships such as the Agroecology Coalition, which can play a key role towards rapidly advancing progress of the Sustainable Development Goals [22,55]. Naturally, addressing long-term and applied research questions will require long-term funding mechanisms from the public and private sectors [52].

For policymakers, our results provide critical evidence of how agricultural diversification can provide long-term support for biodiversity conservation, climate change mitigation, and at least maintaining yields [34]. Biodiversity benefits increase over time without affecting crop yield, despite the commonly expected trade-off associated with biodiversity integration in farmlands [56–58]. The contribution of diversified farming systems to climate change mitigation increases over time, creating win-win solutions between crop yield and climate regulation. Awareness of agricultural diversification is increasing in Europe and North America, but efforts need to be enhanced, particularly in Asia and Africa, to achieve global-scale targets of climate change mitigation and sustainable food production [49]. Agri-environmental schemes and equivalent policies globally should compensate for higher yield gaps in diversified systems to alleviate trade-offs where they exist. Then policies can help restoring degraded and agricultural lands more effectively to reach climate and biodiversity targets [59].

Our work enables farmers to understand short- and long-term benefits of agricultural diversification compared to conventional agriculture or monocultures. A core argument for implementation is a sustainable production strategy that can maintain crop yields and increase financial profitability over time and benefit the environment.. It also enables broader landscape finance strategies, as our results provide detailed evidence of the anticipated trade-offs of agricultural diversification implementation and the associated risks [58]. Agricultural diversification provides further opportunities to shift towards a truly integrated strategy [35], whereby local culture and practices are accounted for, and ecosystem services are fully utilized for diverse livelihoods beyond crop-related productions [60]. In conclusion, our synthesis of the long-term benefits and trade-offs of agricultural diversification globally adds another urgently needed piece of the transformation puzzle for scientists, policymakers, and farmers to make well-informed and urgent decisions for sustainable agricultural systems.

### *Methods*

### *Literature search and data compilation*

We systematically searched for peer-reviewed meta-analysis papers analysing the effects of agricultural diversification on socioeconomic, biological community, soil quality and climate change, using Web of Science ( https://www.webofscience.com/) and Scopus (https://www.scopus.com/). More specifically, we searched for all papers published until December 31, 2022, using keywords related to agricultural diversification practices and socioeconomic and ecological factors related to farming practices, and set meta-analysis as the type of study (for search strings, see Supplementary Table 1). Our research yielded 5,008 papers, which we then screened on Rayyan (https://www.rayyan.ai/) to determine which papers met the criteria for full-text screening. We considered a paper relevant for full-text screening if it was about agricultural production systems, potentially diversified farming systems, and focused on crop yield, financial profitability, biodiversity, pollination, pest control, water regulation, soil fertility, nutrient cycling, carbon sequestration, or climate regulation, and if it was not a primary study. These criteria resulted in 3,692 papers being excluded, leaving 1,316 papers for full-text screening (Supplementary Figure 1).

We screened the full-text of 1,316 papers individually and searched for the following information as inclusion criteria: (1) the study should be a meta-analysis with at least one effect size comparing diversified farming systems to non-diversified practices. (2) The effect size should measure at least one of the following related variables: crop yield, financial profitability, biodiversity, pollination, pest control, water regulation, soil fertility, nutrient cycling, carbon sequestration, and climate regulation (for more details on the variables, see Supplementary Table 2). (3) The number and list of original studies, as well as the number of replicates for each effect size, should be reported. (4) The study should also include information on the duration of diversification practices. In the case of multiple response variables from the same study, we ensured that each response is associated with an independent factor without the possibility of redundancy from one response to another and causing pseudo-replication of the underlying data and effects. These criteria narrowed the list to 192 meta-analyses for data extraction.

We extracted the mean and variance of effect sizes expressed in log response ratio (lnRR), response ratio (RR), and percentage of change, and noted the qualitative effect (i.e., whether it was neutral, significantly positive, or significantly negative). We used Plot Digitizer (http://plotdigitizer.sourceforge.net/ ) to extract data from the figures. We converted the effect sizes from RR and percentage of change into lnRR to obtain a uniform measurement. When the effect size was presented in standardized mean difference (SMD) such as *Hedges' d* or other effect sizes that could not be converted into lnRR without original data, we only extracted the trend (negative or positive) of the effect size as well as the statistical significance, that is, negative, neutral, or positive.

For each effect size, we noted the duration (years) of agricultural diversification practices as well as the number of original comparisons. When the duration of practice was an interval, we considered the median duration (e.g. for 5-10 years, the duration would become 7.5 years). When the duration was indicated as ">x" we added two years, so >10 years would become 12 years to have a single duration for the corresponding effect size. We then categorised each response variable (effect size) into crop yield, biodiversity, pollination, pest control, water regulation, soil fertility, nutrient cycling, carbon sequestration and climate regulation (see Supplementary Table 2). We also categorized agricultural diversification practices into six categories: non-crop diversification, crop diversification, organic amendment, reduced tillage, inoculation, and organic farming (see Supplementary Table 3).

A strong synthesis should not repeat the original studies to guarantee that the effect sizes from all the studies are independent. We avoided including meta-analyses with similar underlying data by first running an independence analysis of 192 studies based on the references of their original studies. We calculated the percentage of original references shared by papers and identified 24 papers with more than 30% of shared original studies. We examined the focus and duration of the effect sizes in these studies. When two or more papers had the same response variable within the same duration, we retained the effect sizes from the most recent study. Using these criteria, we excluded eight papers, resulting in 184 meta-analysis papers with 4,260 effect sizes from 17,989 studies and 169,555 comparisons (Supplementary Figure 1). We evaluated each study's quality based on the methods used and the reported results (for more details, see Supplementary Table 4).

### Data analysis

We synthesized the effects of agricultural diversification on different categories of response variables in overall and per agricultural diversification practice. We used second-order meta-analysis based on hierarchical meta-regression models, weighted with number of comparisons behind each effect size to account for the number of observations behind each effect size. For each model, we used the effect size nested in the study ID (1| StudyID/EffectSizeID) as random effects to account for heterogeneity or between-study variations from the primary studies and effect sizes [61]. First, we ran models predicting the effects of diversification on crop yield, financial profitability, biodiversity, pollination, pest control, water regulation, soil fertility, nutrient cycling, carbon sequestration, and climate regulation individually, without a time moderator (hereafter: moderator), to have reference values for non-temporal syntheses. We then ran the same models by adding the duration of diversification practices as a moderator for both overall and individual agricultural diversification practices. We used the *metafor* r-package [62] to run the meta-regression models, extracted the model results with the r-package orchaRd [63], and used *ggplot2* for visualization [64].

We conducted several tests to detect potential biases in our models. First, we used a funnel plot to visually detect the residuals of the models against the inverse of the standard error (Supplementary Figure 2) with a potential publication bias when the intercept significantly deviated from zero [65]. We ran Egger's regression tests to detect asymmetry of funnel plots (Supplementary Table 5) coupled with fail-safe N analyses based on the Rosenthal method (Supplementary Table 6). We also performed Likelihood

Ratio Tests to detect the statistical significance of publication bias (Supplementary Table 7). We tested the influence of the studies based on the standardized residuals and hat values (Supplementary Figure 3), where the studies with high influence were those with twice the average hat value and with high standardized residuals [66]. Our tests did not reveal any significant issues associated with publication biases or outliers.

To ensure that our results were robust, we performed several sensitivity analyses to compare the outcomes. First, given that we included repeated effect sizes in the response variables, we ran a second-order meta-analysis using a dataset without any repeated responses. Second, since study quality might create some biases, we ran the same analysis but this time with the data set with high-quality studies only. To obtain results from the most conservative analysis, we ran a third type of model with high-quality studies and without repeated responses. These models were run for both the overall effects of diversification and the duration of practices. Overall, the results from these different datasets did not show any significant differences from our main results (i.e., the full dataset, Supplementary Figure 4). For carbon sequestration, excluding repeated variables and/or low-quality studies resulted in neutral long-term effects (>33 years) of diversification, possibly because of data limitations, leading to a wider range of confidence intervals. It is worth noting that all estimates remained positive. Therefore, for coherence of data analyses, we used the full dataset throughout.

We analysed the variation over time in the trade-offs associated with crop yield. We selected studies that tested diversification effects on crop yield and other services at the same time and categorized each effect size as respective, win or lose if the effect size was positive or negative, regardless of their statistical significance. This yielded four standard trade-off outcomes: win_yield - win_Service, win_yield - lose_Service, lose_yield –lose_Service, and lose_yield - win_Service. We then ran hierarchical regression models predicting these outcomes based on a multinomial distribution using the *nnet* r-package [67] and we used the effect size of the crop yield paired with services, nested within the study ID as random effects. We included the mean of the original comparisons of crop yield and paired services as weights to account for the underlying number of comparisons. We ran the models for crop yield with all the effect size categories combined first and then models with the individual categories of effect size categories with the duration of the practices as a moderator.

### Data and code availability

All data and R code used to conduct the analyses and generate graphs in this study are stored in a private online database accessible only to the authors and are available upon request.

### Acknowledgements

We are grateful to Giovanni Tamburini for providing the full texts of the papers used in their studies. E.R. and T.C.W. were funded by the Westlake University start-up fund acquired by T.C.W..

### Author contribution

E.R. and T.C.W. developed the study concept. E.R. collected data, conducted all analyses, produced all figures, and wrote the original draft of the manuscript. T.C.W. secured funding and supervised the project. E. R. and T. C. W. edited and revised the manuscript and approved its submission.

### Reference:

1. Willett, W. *et al.* Food in the Anthropocene: the EAT–Lancet Commission on healthy diets from sustainable food systems. *The Lancet* **393**, 447–492 (2019).
2. Griggs, D. *et al.* Sustainable development goals for people and planet. *Nature* **495**, 305–307 (2013).
3. Kremen, C., Iles, A. & Bacon, C. Diversified farming systems: An agroecological, systems-based alternative to modern industrial agriculture. *Ecology and Society* **17**, (2012).
4. Hodson, E., Niggli, U., Kitajima, K., Lal, R. & Sadoff, C. *Boost Nature Positive Production - A paper on Action Track 3. United Nations Food Systems Summit 2021* https://sc-fss2021.org/wp-content/uploads/2021/04/Action_Track_3_paper_Boost_Nature_Positive_Production.pdf (2021).
5. Tamburini, G. *et al.* Agricultural diversification promotes multiple ecosystem services without compromising yield. *Sci Adv* **6**, (2020).
6. Rosa-Schleich, J., Loos, J., Musshoff, O. & Tscharntke, T. Ecological-economic trade-offs of Diversified Farming Systems - A review. *ECOLOGICAL ECONOMICS* **160**, 251–263 (2019).
7. Gaupp, F. *et al.* Food system development pathways for healthy, nature-positive and inclusive food systems. *Nat Food* **2**, 928–934 (2021).

8. von Braun, J., Afsana, K., Fresco, L., Hassan, M. & Torero, M. Food systems definition, concept and application for the UN food systems summit. United Nations Food Systems Summit 2021 (2021).

9. FAO. The 10 elements of Agroecology Guiding the transition to sustainable food and agricultural systems. *Fao* 15 (2019).

10. HLPE. Food security and nutrition: building a global narrative towards 2030. A report by the High Level Panel of Experts on Food Security and Nutrition of the Committee on World Food Security, Rome. *High Level Panel of Experts* 112 (2020).

11. Wezel, A. *et al.* Agroecological principles and elements and their implications for transitioning to sustainable food systems. A review. *Agron Sustain Dev* **40**, 40 (2020).

12. Gurr, G. M. *et al.* Multi-country evidence that crop diversification promotes ecological intensification of agriculture. *Nat Plants* **2**, 16014 (2016).

13. He, X. *et al.* Integrating agricultural diversification in China's major policies. *Trends Ecol Evol* **37**, 819–822 (2022).

14. Smith, O. M. *et al.* Organic Farming Provides Reliable Environmental Benefits but Increases Variability in Crop Yields: A Global Meta-Analysis. *Front Sustain Food Syst* **3**, (2019).

15. Muneret, L. *et al.* Evidence that organic farming promotes pest control. *Nat Sustain* **1**, 361–368 (2018).

16. Gattinger, A. *et al.* Enhanced top soil carbon stocks under organic farming. *Proceedings of the National Academy of Sciences* **109**, 18226–18231 (2012).

17. Marrec, R., Brusse, T. & Caro, G. Biodiversity-friendly agricultural landscapes – integrating farming practices and spatiotemporal dynamics. *Trends Ecol Evol* **37**, 731–733 (2022).

18. Tscharntke, T., Grass, I., Wanger, T. C., Westphal, C. & Batáry, P. Spatiotemporal land-use diversification for biodiversity. *Trends Ecol Evol* **37**, 734–735 (2022).

19. Klapwijk, C. J. *et al.* Analysis of trade-offs in agricultural systems: Current status and way forward. *Current Opinion in Environmental Sustainability* vol. 6 110–115 Preprint at https://doi.org/10.1016/j.cosust.2013.11.012 (2014).

20. He, X. *et al.* Agricultural diversification promotes sustainable and resilient global rice production. *Nat Food* **4**, 788–796 (2023).

21. Toledo-Hernández, M. *et al.* Genome-edited tree crops: mind the socioeconomic implementation gap. *Trends Ecol Evol* **36**, 972–975 (2021).

22. Niggli, U., Sonnevelt, M. & Kummer, S. Pathways to Advance Agroecology for a Successful Transformation to Sustainable Food Systems. *Science and Innovations for Food Systems Transformation* 341–359 (2023) doi:10.1007/978-3-031-15703-5_18.

23. Mehrabi, Z. *et al.* Research priorities for global food security under extreme events. *One Earth* **5**, 756–766 (2022).

24. Ruckelshaus, M. H. *et al.* The IPBES Global Assessment: Pathways to Action. *Trends Ecol Evol* **35**, 407–414 (2020).

25. IPBES. Summary for policymakers of the global assessment report on biodiversity and ecosystem services of the Intergovernmental Science-Policy Platform on Biodiversity and Ecosystem Services. (2019).

26. Qi, G. *et al.* Yield Responses of Wheat to Crop Residue Returning in China: A Meta-Analysis. *Crop Sci* **59**, 2185–2200 (2019).

27. Renard, D. & Tilman, D. National food production stabilized by crop diversity. *Nature* **571**, 257–260 (2019).

28. WWF. Bending the curve of biodiversity loss. Living Planet Report 2020 (2020).

29. Dainese, M. *et al.* A global synthesis reveals biodiversity-mediated benefits for crop production. *Sci Adv* **5**, 1–14 (2019).

30. Klein, A.-M. *et al.* Importance of pollinators in changing landscapes for world crops. *Proceedings of the Royal Society B: Biological Sciences* **274**, 303–313 (2007).

31. Tscharntke, T. *et al.* When natural habitat fails to enhance biological pest control – Five hypotheses. *Biol Conserv* **204**, 449–458 (2016).

32. Bai, X. *et al.* Responses of soil carbon sequestration to climate-smart agriculture practices: A meta-analysis. *Glob Chang Biol* **25**, 2591–2606 (2019).

33. Lessmann, M., Ros, G. H., Young, M. D. & Vries, W. Global variation in soil carbon sequestration potential through improved cropland management. *Glob Chang Biol* **28**, 1162–1177 (2022).

34. Searchinger, T. *et al.* Creating a sustainable food future: A menu of solutions to feed more than 9 billion people by 2050. *World Resources Institute* 1–5 (2014).

35. Leclère, D. *et al.* Bending the curve of terrestrial biodiversity needs an integrated strategy. *Nature* **585**, 551–556 (2020).

36. Qin, Z., Dunn, J. B., Kwon, H., Mueller, S. & Wander, M. M. Soil carbon sequestration and land use change associated with biofuel production: empirical evidence. *GCB Bioenergy* **8**, 66–80 (2016).

37. Guenet, B. *et al.* Can N2O emissions offset the benefits from soil organic carbon storage? *Glob Chang Biol* **27**, 237–256 (2021).

38. Chen, Y., Camps-Arbestain, M., Shen, Q., Singh, B. & Cayuela, M. L. The long-term role of organic amendments in building soil nutrient fertility: a meta-analysis and review. *Nutr Cycl Agroecosyst* **111**, 103–125 (2018).

39. Ledo, A. *et al.* Changes in soil organic carbon under perennial crops. *Glob Chang Biol* **26**, 4158–4168 (2020).

40. Wei, Z., Hoffland, E., Zhuang, M., Hellegers, P. & Cui, Z. Organic inputs to reduce nitrogen export via leaching and runoff: A global meta-analysis. *Environmental Pollution* **291**, 118176 (2021).

41. Tesfaye, F. *et al.* Could biochar amendment be a tool to improve soil availability and plant uptake of phosphorus? A meta-analysis of published experiments. *Environmental Science and Pollution Research* **28**, 34108–34120 (2021).

42. Sánchez, A. C., Kamau, H. N., Grazioli, F. & Jones, S. K. Financial profitability of diversified farming systems: A global meta-analysis. *Ecological Economics* **201**, 107595 (2022).

43. Smith, M. E. *et al.* Increasing crop rotational diversity can enhance cereal yields. *Commun Earth Environ* **4**, 89 (2023).

44. Pumariño, L. *et al.* Effects of agroforestry on pest, disease and weed control: A meta-analysis. *Basic Appl Ecol* **16**, 573–582 (2015).

45. Niether, W., Jacobi, J., Blaser, W. J., Andres, C. & Armengot, L. Cocoa agroforestry systems versus monocultures: a multi-dimensional meta-analysis. *Environmental Research Letters* **15**, 104085 (2020).

46. Tscharntke, T., Grass, I., Wanger, T. C., Westphal, C. & Batáry, P. Beyond organic farming – harnessing biodiversity-friendly landscapes. *Trends Ecol Evol* **36**, 919–930 (2021).

47. Cui, Z. *et al.* Pursuing sustainable productivity with millions of smallholder farmers. *Nature* **555**, 363–366 (2018).

48. Hertel, T. W., Elouafi, I., Ewert, F. & Tanticharoen, morakot. Building Resilience to Vulnerabilities, Shocks and Stresses. *UN Food Systems Summit* 1–24 (2021).

49. FAO, European Union & CIRAD. Catalysing the sustainable and inclusive transformation of food systems. (2021).

50. Egli, L., Meyer, C., Scherber, C., Kreft, H. & Tscharntke, T. Winners and losers of national and global efforts to reconcile agricultural intensification and biodiversity conservation. *Glob Chang Biol* **24**, 2212–2228 (2018).

51. Martinez-Salinas, A. *et al.* Interacting pest control and pollination services in coffee systems. *Proc Natl Acad Sci U S A* **119**, 1–7 (2022).

52. Johnston, A. E. & Poulton, P. R. The importance of long-term experiments in agriculture: their management to ensure continued crop production and soil fertility; the Rothamsted experience. *Eur J Soil Sci* **69**, 113–125 (2018).

53. Ortiz-Bobea, A., Ault, T. R., Carrillo, C. M., Chambers, R. G. & Lobell, D. B. Anthropogenic climate change has slowed global agricultural productivity growth. *Nat Clim Chang* **11**, 306–312 (2021).

54. Aggarwal, P., Vyas, S., Thornton, P. & Campbell, B. M. How much does climate change add to the challenge of feeding the planet this century? *Environmental Research Letters* **14**, (2019).

55. Montenegro, M. & Wit, D. Operating principles for collective scholar-activism : Early insights from the Agroecology Research-Action Collective. **10**, 319–337 (2021).

56. Holt, A. R., Alix, A., Thompson, A. & Maltby, L. Science of the Total Environment Food production , ecosystem services and biodiversity : We can ' t have it all everywhere. *Science of the Total Environment, The* **573**, 1422–1429 (2016).

57. Hegwood, M., Langendorf, R. E. & Burgess, M. G. Why win–wins are rare in complex environmental management. *Nat Sustain* **5**, 674–680 (2022).

58. Kuyah, S. *et al.* Agroforestry delivers a win-win solution for ecosystem services in sub-Saharan Africa. A meta-analysis. *Agron Sustain Dev* **39**, (2019).

59. Mo, L. *et al.* Integrated global assessment of the natural forest carbon potential. *Nature* **624**, (2023).

60. IPBES. Methodological Assessment Report on the Diverse Values and Valuation of Nature of the Intergovernmental Science-Policy Platform on Biodiversity and Ecosystem Services. (IPBES secretariat, Bonn, Germany., 2022). doi:https://doi.org/10.5281/zenodo.6522522.

61. Schmidt, F. L. & Oh, I. S. Methods for second order meta-analysis and illustrative applications. *Organ Behav Hum Decis Process* **121**, 204–218 (2013).

62. Viechtbauer, W. Conducting meta-analyses in R with the metafor. *J Stat Softw* **36**, 1–48 (2010).

63. Nakagawa, S. *et al.* orchaRd 2.0: An R package for visualising meta-analyses with orchard plots. *Methods Ecol Evol* **2023**, 1–8 (2023).

64. Wickham, H. *ggplot2: Elegant Graphics for Data Analysis.* (Springer-Verlag New York, 2016).

65. Viechtbauer, W. Model Checking in Meta-Analysis. in *Handbook of Meta-Analysis* (eds. Schmid, C. H., Stijnen, T. & White, I.) 219–254 (Chapman and Hall/CRC, 2020).

66. Marks-Anglin, A., Duan, R., Chen, Y., Panagiotou, O. & Schmid, C. H. Publication and Outcome Reporting Bias. in *Handbook of Meta-Analysis* (eds. Schmid, C. H., Stijnen, T. & White, I. R.) (Chapman and Hall/CRC, 2020).

67. Brooks, M. E. *et al.* glmmTMB Balances Speed and Flexibility Among Packages for Zero-inflated Generalized Linear Mixed Modeling. *R J* **9**, 378–400 (2017).

**_Title:_** **_Decades matter: Agricultural diversification increases financial profitability, biodiversity, and ecosystem services over time._**

Authors: Estelle Raveloaritiana[1,2,*,] Thomas Cherico Wanger [1,2,3,4,5*]

[1] Sustainable Agricultural Systems & Engineering Laboratory, School of Engineering, Westlake University, Hangzhou, China.
[2] Key Laboratory of Coastal Environment and Resources of Zhejiang Province, Westlake University, Hangzhou, China.
[3] Faculty of Science, Technology & Medicine, University of Luxembourg, Luxembourg
[4] GlobalAgroforestryNetwork.org, Hangzhou, China
[5] ChinaRiceNetwork.org, Hangzhou, China

*Corresponding authors: eraveloaritiana@gmail.com (ER); tomcwanger@gmail.com (TCW)

# Extended Data and Supplementary materials

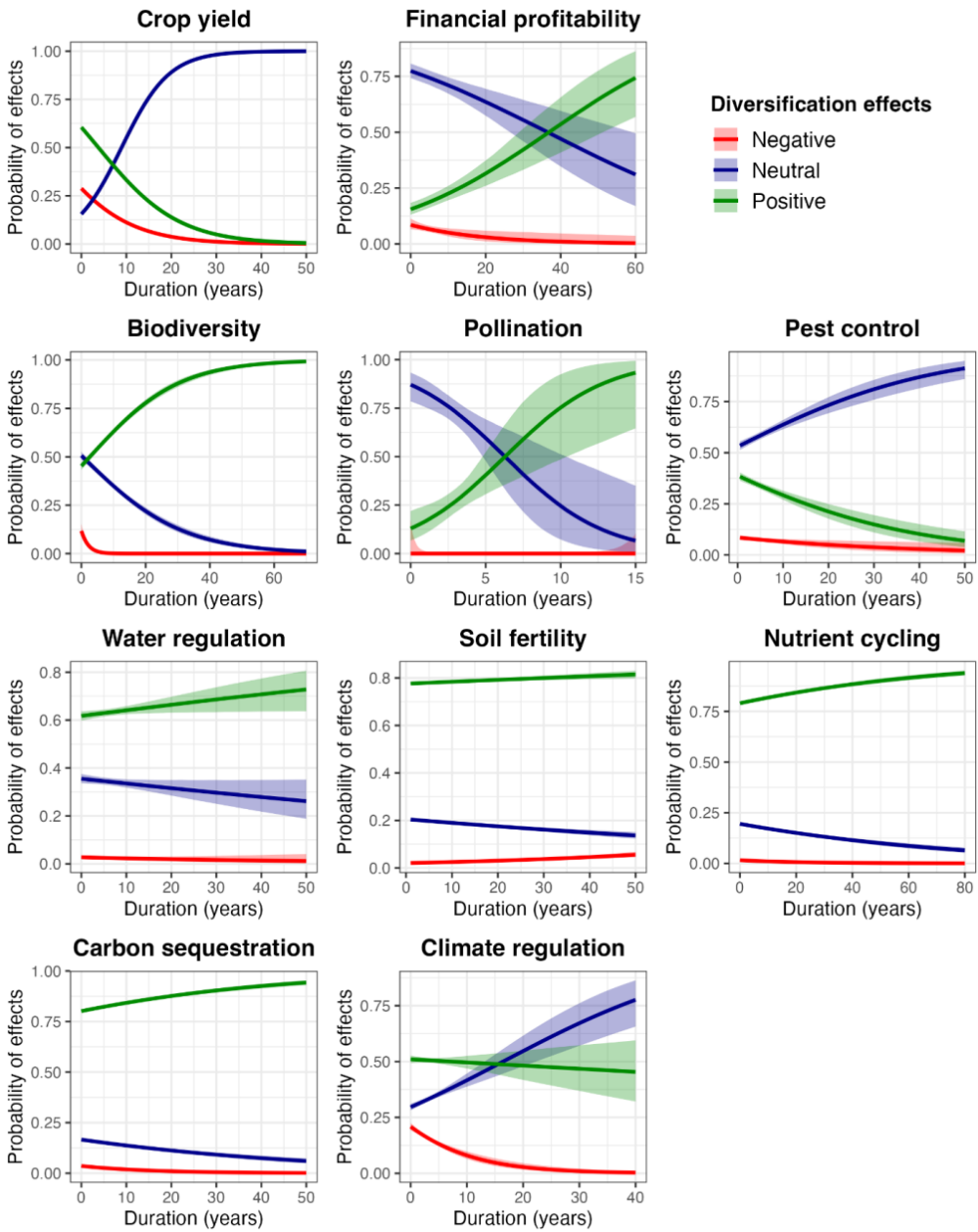

Extended Data Fig. 1| Quantitative systematic review of the effects of the diversification practices on finance, biodiversity, and ecosystem services

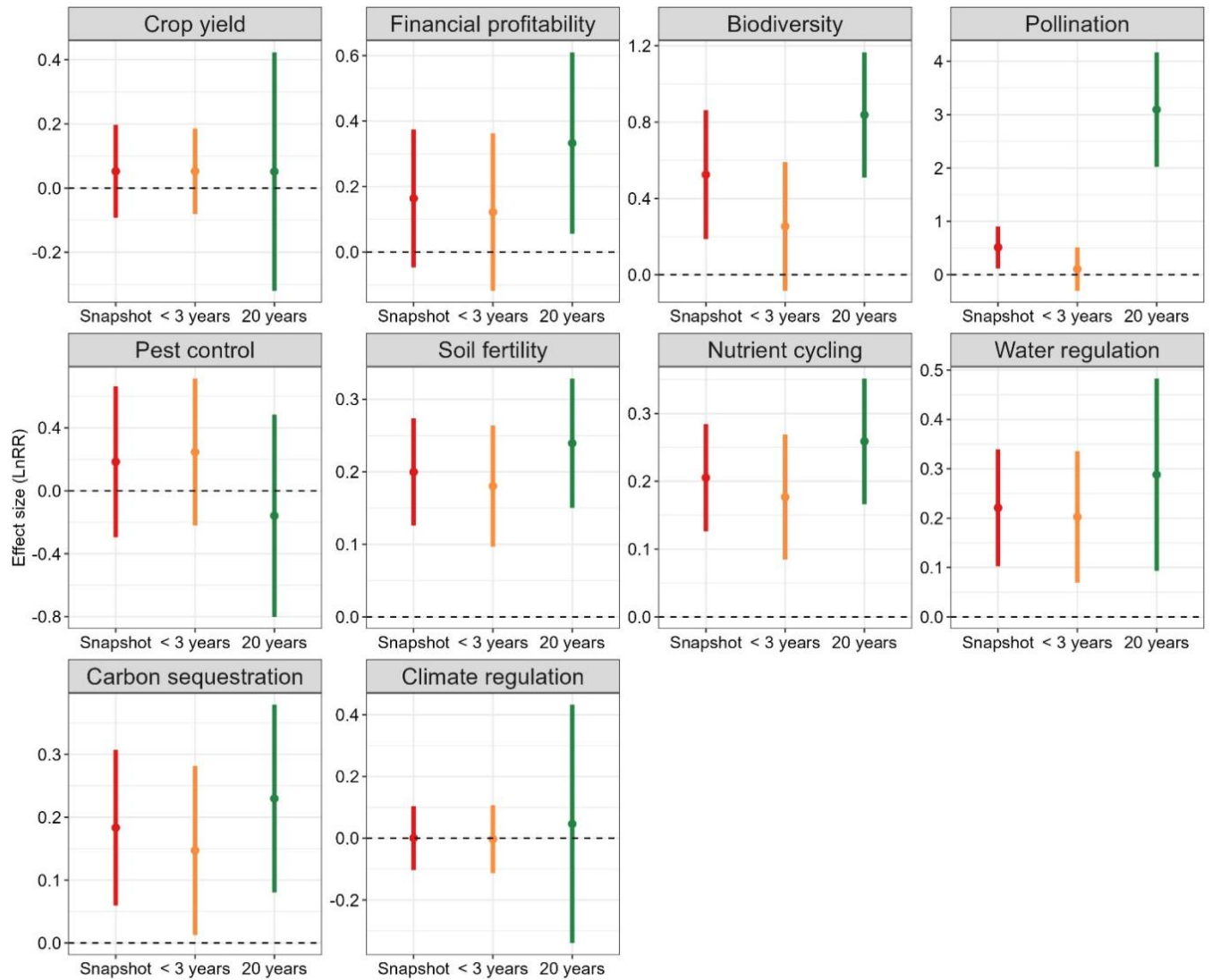

Extended Data Fig. 2| Snapshot, short-term (<3 years) and long-term (20 years) effects of diversification practices on socioeconomic, biological community, soil quality, and climate change mitigation related variables

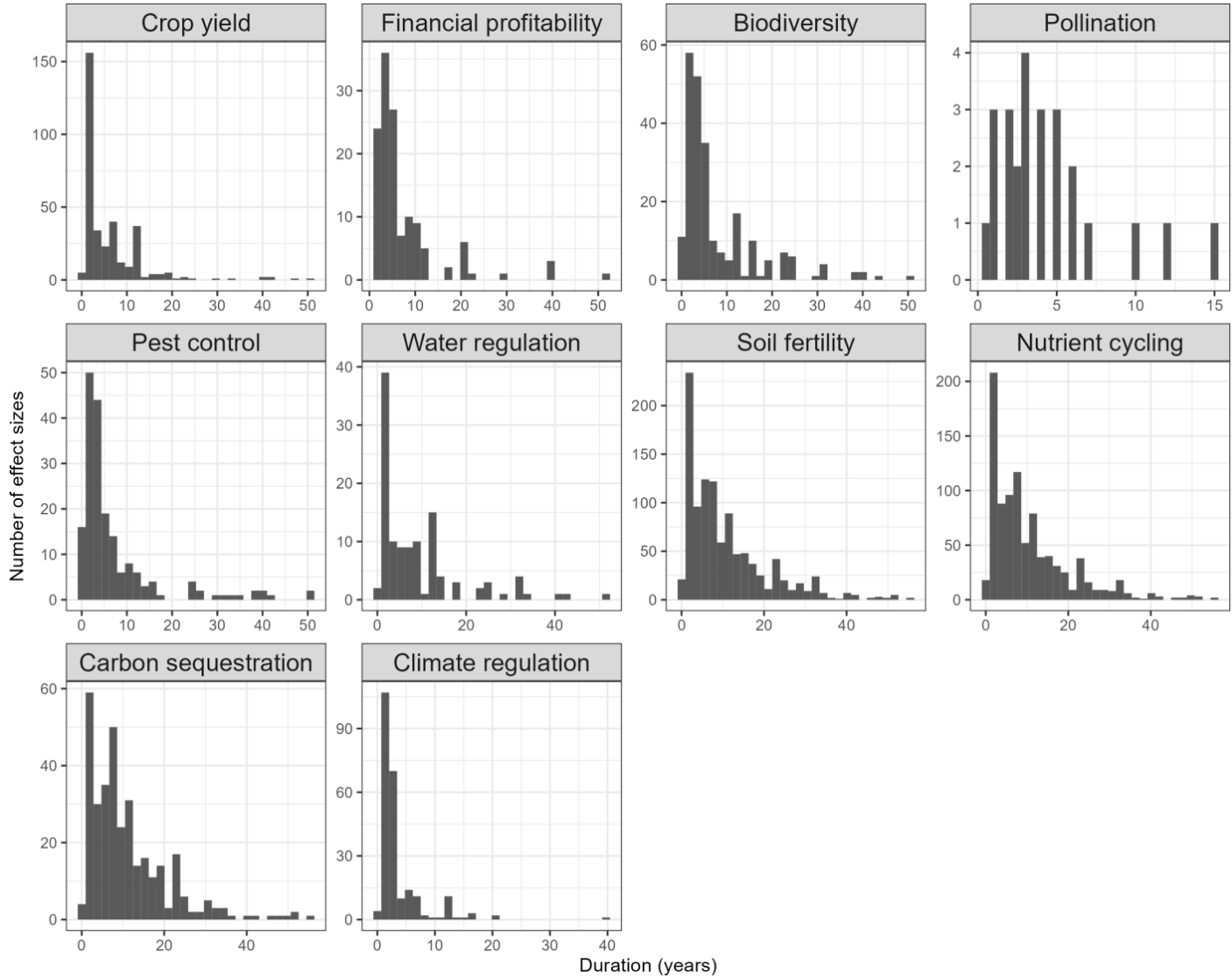

Extended Data Fig. 3| Vote count of the effect sizes included in the effects of agricultural diversification over time

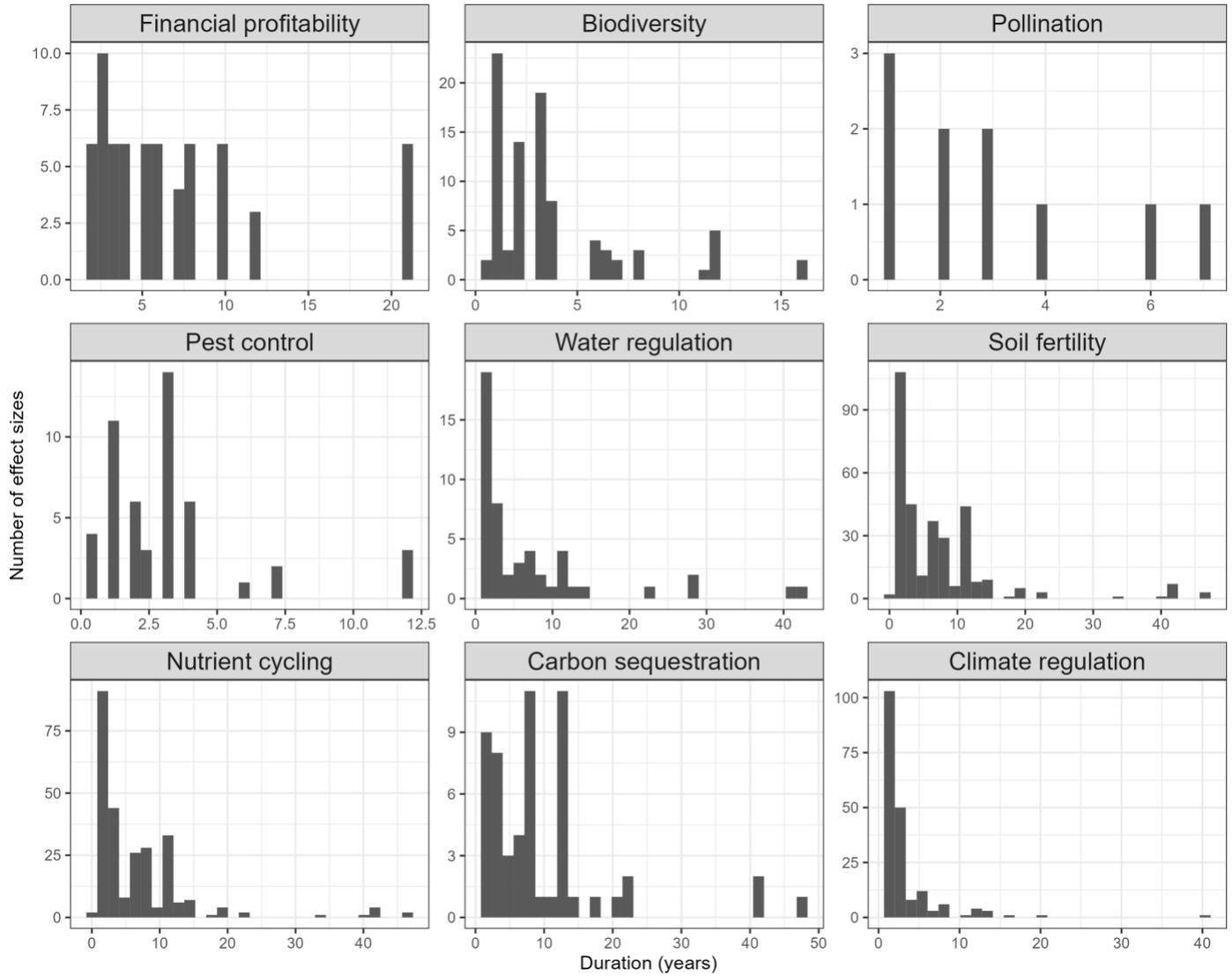

Extended Data Fig. 4| Vote count of the effect sizes included in the trade-off analysis between crop yield and other services across the duration of the diversification practices.

Extended Data Table 1. Estimates, standard errors, p-values and r-squared of the response variables depict the effects of overall diversification practices without and with duration as moderator.

| No moderators | | | | | | |
|---|---|---|---|---|---|---|
| | Moderators | Estimate | SE | z-value | p-value | |
| Crop yield | - | 0.0524 | 0.0737 | 0.7108 | 0.4772 | |
| Financial profitability | - | 0.1639 | 0.1074 | 1.5258 | 0.1271 | |
| Biodiversity | - | 0.5247 | 0.1722 | 3.0465 | 0.0023 | |
| Pollination | - | 0.5125 | 0.2005 | 2.5553 | 0.0106 | |
| Pest control | - | 0.1845 | 0.2449 | 0.7532 | 0.4513 | |
| Soil fertility | - | 0.1998 | 0.0377 | 5.2938 | <.0001 | |
| Nutrient cycling | - | 0.2053 | 0.0403 | 5.0944 | <.0001 | |
| Water regulation | - | 0.221 | 0.0604 | 3.6617 | 0.0003 | |
| Carbon sequestration | - | 0.1836 | 0.0631 | 2.9096 | 0.0036 | |
| Climate regulation | - | 0.0004 | 0.0528 | 0.008 | 0.9936 | |
| With Duration as moderator | | | | | | |
| Response variables | Moderators | Estimate | SE | z-value | p-value | Marginal $R^2$ |
| Crop yield | Intercept | 0.0526 | 0.0668 | 0.7863 | 0.4317 | 0.000 |
| | Duration (years) | 0 | 0.0089 | -0.0053 | 0.9957 | |
| Financial profitability | Intercept | 0.1024 | 0.1245 | 0.8228 | 0.4106 | 0.128 |
| | Duration (years) | 0.0113 | 0.0052 | 2.1796 | 0.0293 | |
| Biodiversity | Intercept | 0.1984 | 0.1794 | 1.1062 | 0.2687 | 0.204 |
| | Duration (years) | 0.031 | 0.0083 | 3.7512 | 0.0002 | |
| Pollination | Intercept | -0.6172 | 0.2781 | -2.2195 | 0.0265 | 0.981 |
| | Duration (years) | 0.3548 | 0.068 | 5.2171 | <.0001 | |
| Pest control | Intercept | 0.2947 | 0.2339 | 1.2597 | 0.2078 | 0.146 |
| | Duration (years) | -0.022 | 0.0093 | -2.3623 | 0.0182 | |
| Soil fertility | Intercept | 0.1739 | 0.0452 | 3.8486 | 0.0001 | 0.017 |
| | Duration (years) | 0.0032 | 0.0025 | 1.2884 | 0.1976 | |
| Nutrient cycling | Intercept | 0.1679 | 0.0498 | 3.3716 | 0.0007 | 0.030 |
| | Duration (years) | 0.0044 | 0.0026 | 1.7261 | 0.0843 | |
| Water regulation | Intercept | 0.1944 | 0.0735 | 2.6456 | 0.0082 | 0.035 |
| | Duration (years) | 0.0046 | 0.0059 | 0.7689 | 0.4419 | |
| Carbon sequestration | Intercept | 0.1384 | 0.0716 | 1.934 | 0.031 | 0.025 |
| | Duration (years) | 0.0045 | 0.0036 | 1.2395 | 0.2151 | |
| Climate regulation | Intercept | -0.0079 | 0.0641 | -0.1227 | 0.9023 | 0.002 |
| | Duration (years) | 0.0026 | 0.0109 | 0.241 | 0.8095 | |

Extended Data Table 2. Estimates of non-temporal effects sizes as well as the short term (<3 years) and long-term (20 years) estimates and the percentage differences amongst the estimates.

| | Non-temporal Estimate | Estimate 0 to 3 years | estimate 20 years | Percentage difference between non-temporal and < 3 years | Percentage difference between non-temporal and 20 years estimates | Percentage difference between < 3 years and 20 years estimates |
|---|---|---|---|---|---|---|
| Crop yield | 0.0524 | 0.0525 | 0.0526 | 0 | 0 | 0 |
| Financial profitability | 0.1639 | 0.1222 | 0.3327 | -25 | 100 | 172 |
| Biodiversity | 0.5247 | 0.2534 | 0.8377 | -52 | 56 | 231 |
| Pollination** | 0.5125 | 0.1059 | 3.0939 | -79 | 1164 | 2823 |
| Pest control | 0.1845 | 0.2463 | -0.1588 | 33 | -179 | -164 |
| Soil fertility | 0.1998 | 0.1803 | 0.2394 | -10 | 19 | 33 |
| Nutrient cycling | 0.2053 | 0.1768 | 0.2589 | -14 | 25 | 46 |
| Water regulation | 0.2210 | 0.2026 | 0.2881 | -8 | 30 | 42 |
| Carbon sequestration | 0.1836 | 0.1474 | 0.2300 | -20 | 24 | 56 |
| Climate regulation | 0.0004 | 0.0001 | 0.0462 | -75 | 10925 | 46103 |

**10 years for Pollination

Extended Data Table 3. Estimates, standard errors, and p-values of the response variables showing the effects of different categories of diversification practices on all response variables without temporal consideration.

| Crop diversification | | | | |
|---|---|---|---|---|
| | Estimate | SE | z-value | p-value |
| Crop yield | 0.0365 | 0.0832 | 0.4391 | 0.6606 |
| Financial profitability | 0.1548 | 0.114 | 1.3579 | 0.1745 |
| Biodiversity | 0.2427 | 0.2954 | 0.8216 | 0.4113 |
| Pollination | 0.1816 | 0.1532 | 1.1852 | 0.2359 |
| Pest control | 0.2848 | 0.2792 | 1.0199 | 0.3078 |
| Water regulation | 0.5619 | 0.4344 | 1.2935 | 0.1958 |
| Soil fertility | 0.1731 | 0.0706 | 2.4521 | 0.0142 |
| Nutrient cycling | 0.1932 | 0.075 | 2.5768 | 0.01 |
| Carbon sequestration | 0.1558 | 0.1323 | 1.1777 | 0.2389 |
| Climate regulation | 0.2238 | 0.2112 | 1.0597 | 0.2893 |
| Non-crop diversification | | | | |
| | Estimate | SE | z-value | p-value |
| Crop yield | 0.2783 | 0.149 | 1.8679 | 0.0618 |
| Financial profitability | -0.0335 | 0.4925 | -0.068 | 0.9458 |
| Biodiversity | 0.5212 | 0.0722 | 7.2203 | <.0001 |

| Pollination | 1.1824 | 0.373 | 3.1705 | 0.0015 |
|---|---|---|---|---|
| Pest control | 0.1851 | 0.0518 | 3.5709 | 0.0004 |
| Water regulation | 0.477 | 0.2618 | 1.8223 | 0.0684 |
| Soil fertility | 0.2233 | 0.078 | 2.8614 | 0.0042 |
| Nutrient cycling | 0.2569 | 0.1176 | 2.184 | 0.029 |
| Carbon sequestration | 0.3652 | 0.0795 | 4.5923 | <.0001 |
| Climate regulation | -0.0756 | 0.0863 | -0.8768 | 0.3806 |
| Reduced tillage | | | | |
| | Estimate | SE | z-value | p-value |
| Crop yield | -0.0276 | 0.0232 | -1.1893 | 0.2343 |
| Biodiversity | 0.4037 | 0.1904 | 2.1199 | 0.034 |
| Water regulation | 0.0896 | 0.1042 | 0.8601 | 0.3898 |
| Soil fertility | 0.1539 | 0.062 | 2.4812 | 0.0131 |
| Nutrient cycling | 0.1402 | 0.0578 | 2.4274 | 0.0152 |
| Carbon sequestration | 0.1014 | 0.0108 | 9.3981 | <.0001 |
| Climate regulation | 0.0236 | 0.0997 | 0.2366 | 0.8129 |
| Organic amendment | | | | |
| | Estimate | SE | z-value | p-value |
| Crop yield | 0.1105 | 0.0287 | 3.8512 | 0.0001 |
| Biodiversity | 0.7264 | 0.1916 | 3.7905 | 0.0002 |
| Water regulation | 0.2151 | 0.0632 | 3.4023 | 0.0007 |
| Soil fertility | 0.2139 | 0.0702 | 3.0488 | 0.0023 |
| Nutrient cycling | 0.223 | 0.0638 | 3.4951 | 0.0005 |
| Carbon sequestration | 0.2013 | 0.0874 | 2.3022 | 0.0213 |
| Climate regulation | -0.0114 | 0.0773 | -0.1476 | 0.8827 |
| Organic farming | | | | |
| | Estimate | SE | z-value | p-value |
| Crop yield | -0.2027 | 0.1719 | -1.1788 | 0.2385 |
| Financial profitability | 0.1821 | 0.1477 | 1.2329 | 0.2176 |
| Biodiversity | 0.612 | 0.578 | 1.0589 | 0.2897 |
| Pollination | 0.1959 | 0.0658 | 2.9751 | 0.0029 |
| Pest control | -0.0685 | 0.1946 | -0.3519 | 0.7249 |
| Water regulation | 0.27 | 0.22 | 1.2274 | 0.2197 |
| Soil fertility | 0.3866 | 0.1421 | 2.7199 | 0.0065 |
| Nutrient cycling | 0.3847 | 0.177 | 2.1738 | 0.0297 |
| Carbon sequestration | 0.8812 | 0.5559 | 1.5853 | 0.1129 |
| Inoculation | | | | |
| | Estimate | SE | z-value | p-value |
| Crop yield | 0.1533 | 0.1782 | 0.8602 | 0.3897 |
| Soil fertility | 0.3382 | 0.1605 | 2.107 | 0.0351 |
| Nutrient cycling | 0.1593 | 0.0601 | 2.651 | 0.008 |
| Climate regulation | 0.0536 | 0.1153 | 0.4647 | 0.6421 |

Extended Data Table 4. Estimates, standard errors, p-values and r-squared of the response variables showing the effects of <u>different categories of diversification practices</u> on all response variables over time.

| Crop diversification | | | | | | |
|---|---|---|---|---|---|---|
| **Response variables** | Moderators | Estimate | SE | z-value | p-value | Marginal $R^2$ |
| **Crop yield** | Intercept | 0.037 | 0.103 | 0.361 | 0.718 | 0.000 |
| | Duration (years) | 0.000 | 0.009 | -0.023 | 0.982 | |
| **Financial profitability** | Intercept | 0.090 | 0.154 | 0.583 | 0.560 | 0.281 |
| | Duration (years) | 0.011 | 0.005 | 2.477 | 0.013 | |
| **Biodiversity** | Intercept | 0.165 | 0.298 | 0.554 | 0.580 | 0.083 |
| | Duration (years) | 0.009 | 0.008 | 1.089 | 0.276 | |
| **Pollination** | Intercept | 0.527 | 0.736 | 0.716 | 0.474 | 0.111 |
| | Duration (years) | -0.104 | 0.119 | -0.874 | 0.382 | |
| **Pest control** | Intercept | 0.408 | 0.270 | 1.510 | 0.131 | 0.224 |
| | Duration (years) | -0.024 | 0.011 | -2.313 | 0.021 | |
| **Soil fertility** | Intercept | 0.201 | 0.091 | 2.217 | 0.027 | 0.007 |
| | Duration (years) | -0.002 | 0.004 | -0.463 | 0.644 | |
| **Nutrient cycling** | Intercept | 0.228 | 0.098 | 2.322 | 0.020 | 0.008 |
| | Duration (years) | -0.002 | 0.004 | -0.550 | 0.582 | |
| **Water regulation** | Intercept | 0.897 | 0.605 | 1.482 | 0.138 | 0.276 |
| | Duration (years) | -0.074 | 0.076 | -0.974 | 0.330 | |
| **Carbon sequestration** | Intercept | 0.187 | 0.170 | 1.103 | 0.270 | 0.005 |
| | Duration (years) | -0.002 | 0.008 | -0.282 | 0.778 | |
| **Climate regulation** | Intercept | 0.113 | 0.456 | 0.248 | 0.804 | 0.010 |
| | Duration (years) | 0.086 | 0.285 | 0.301 | 0.764 | |
| **Non-crop diversification** | | | | | | |
| **Response variables** | Moderators | Estimate | SE | z-value | p-value | Marginal $R^2$ |
| **Crop yield** | Intercept | -0.395 | 0.254 | -1.555 | 0.120 | 0.668 |
| | Duration (years) | 0.132 | 0.051 | 2.616 | 0.009 | |
| **Biodiversity** | Intercept | 0.359 | 0.153 | 2.344 | 0.019 | 0.547 |
| | Duration (years) | 0.043 | 0.013 | 3.411 | 0.001 | |
| **Pollination** | Intercept | -0.271 | 0.513 | -0.527 | 0.598 | 0.998 |
| | Duration (years) | 0.401 | 0.083 | 4.851 | <.0001 | |
| **Pest control** | Intercept | 0.038 | 0.093 | 0.405 | 0.685 | 0.314 |
| | Duration (years) | 0.050 | 0.030 | 1.647 | 0.100 | |
| **Soil fertility** | Intercept | 0.070 | 0.094 | 0.748 | 0.454 | 0.486 |
| | Duration (years) | 0.014 | 0.005 | 2.740 | 0.006 | |
| **Nutrient cycling** | Intercept | 0.081 | 0.180 | 0.451 | 0.652 | 0.287 |
| | Duration (years) | 0.014 | 0.009 | 1.650 | 0.099 | |
| **Water regulation** | Intercept | 0.4144 | 0.2651 | 1.5634 | 0.118 | 0.018 |
| | Duration (years) | 0.0036 | 0.0103 | 0.3543 | 0.7231 | |
| **Carbon sequestration** | Intercept | 0.0928 | 0.043 | 2.16 | 0.0308 | 0.896 |
| | Duration (years) | 0.0156 | 0.0024 | 6.5079 | <.0001 | |
| **Reduced tillage** | | | | | | |
| **Response variables** | Moderators | Estimate | SE | z-value | p-value | Marginal $R^2$ |
| **Crop yield** | Intercept | -0.039 | 0.024 | -1.639 | 0.101 | 0.032 |

| | | | | | |
|---|---|---|---|---|---|
| | Duration (years) | 0.002 | 0.002 | 1.112 | 0.266 | |
| **Biodiversity** | Intercept | 0.327 | 0.256 | 1.280 | 0.201 | 0.032 |
| | Duration (years) | 0.008 | 0.011 | 0.671 | 0.503 | |
| **Soil fertility** | Intercept | 0.165 | 0.059 | 2.811 | 0.005 | 0.002 |
| | Duration (years) | -0.001 | 0.003 | -0.319 | 0.750 | |
| **Nutrient cycling** | Intercept | 0.126 | 0.047 | 2.690 | 0.007 | 0.004 |
| | Duration (years) | 0.001 | 0.003 | 0.410 | 0.682 | |
| **Water regulation** | Intercept | 0.103 | 0.105 | 0.980 | 0.327 | 0.006 |
| | Duration (years) | -0.002 | 0.009 | -0.237 | 0.813 | |
| **Carbon sequestration** | Intercept | 0.077 | 0.015 | 5.115 | <.0001 | 0.580 |
| | Duration (years) | 0.002 | 0.001 | 2.154 | 0.031 | |
| **Climate regulation** | Intercept | 0.056 | 0.098 | 0.574 | 0.566 | 0.006 |
| | Duration (years) | -0.006 | 0.011 | -0.518 | 0.605 | |

**Organic amendment**

| Response variables | Moderators | Estimate | SE | z-value | p-value | Marginal $R^2$ |
|---|---|---|---|---|---|---|
| **Crop yield** | Intercept | 0.077 | 0.035 | 2.235 | 0.025 | 0.344 |
| | Duration (years) | 0.010 | 0.006 | 1.702 | 0.089 | |
| **Biodiversity** | Intercept | 0.141 | 0.095 | 1.482 | 0.138 | 0.891 |
| | Duration (years) | 0.041 | 0.004 | 10.719 | <.0001 | |
| **Soil fertility** | Intercept | 0.185 | 0.079 | 2.338 | 0.019 | 0.036 |
| | Duration (years) | 0.005 | 0.005 | 1.168 | 0.243 | |
| **Nutrient cycling** | Intercept | 0.165 | 0.072 | 2.279 | 0.023 | 0.142 |
| | Duration (years) | 0.010 | 0.004 | 2.668 | 0.008 | |
| **Water regulation** | Intercept | 0.221 | 0.069 | 3.211 | 0.001 | 0.022 |
| | Duration (years) | -0.002 | 0.008 | -0.247 | 0.805 | |
| **Carbon sequestration** | Intercept | 0.166 | 0.104 | 1.593 | 0.111 | 0.035 |
| | Duration (years) | 0.006 | 0.007 | 0.916 | 0.360 | |
| **Climate regulation** | Intercept | -0.030 | 0.085 | -0.353 | 0.724 | 0.012 |
| | Duration (years) | 0.007 | 0.018 | 0.392 | 0.695 | |

**Organic farming**

| Response variables | Moderators | Estimate | SE | z-value | p-value | Marginal $R^2$ |
|---|---|---|---|---|---|---|
| **Crop yield** | Intercept | -0.223 | 0.225 | -0.987 | 0.324 | 0.008 |
| | Duration (years) | 0.003 | 0.012 | 0.234 | 0.815 | |
| **Financial profitability** | Intercept | 0.116 | 0.196 | 0.592 | 0.554 | 0.044 |
| | Duration (years) | 0.014 | 0.024 | 0.577 | 0.564 | |
| **Biodiversity** | Intercept | 0.037 | 0.689 | 0.053 | 0.958 | 0.064 |
| | Duration (years) | 0.039 | 0.026 | 1.509 | 0.131 | |
| **Pollination** | Intercept | 0.208 | 0.259 | 0.805 | 0.421 | 0.009 |
| | Duration (years) | -0.006 | 0.099 | -0.055 | 0.956 | |
| **Pest control** | Intercept | 0.023 | 0.184 | 0.124 | 0.901 | 0.237 |
| | Duration (years) | -0.016 | 0.016 | -1.043 | 0.297 | |
| **Soil fertility** | Intercept | 0.4611 | 0.2507 | 1.8397 | 0.0658 | 0.029 |
| | Duration (years) | -0.0064 | 0.0156 | -0.414 | 0.6789 | |
| **Nutrient cycling** | Intercept | 0.394 | 0.250 | 1.575 | 0.115 | 0.0003 |
| | Duration (years) | -0.001 | 0.020 | -0.052 | 0.959 | |
| **Carbon sequestration** | Intercept | 1.775 | 0.317 | 5.600 | <.0001 | 0.87 |
| | Duration (years) | -0.050 | 0.026 | -1.906 | 0.057 | |

| Inoculation | | | | | | |
|---|---|---|---|---|---|---|
| **Response variables** | Moderators | Estimate | SE | z-value | p-value | Marginal $R^2$ |
| **Crop yield** | Intercept | 0.067 | 0.228 | 0.293 | 0.769 | 0.031 |
| | Duration (years) | 0.051 | 0.060 | 0.855 | 0.392 | |
| **Soil fertility** | Intercept | 0.617 | 0.225 | 2.747 | 0.006 | 0.485 |
| | Duration (years) | -0.187 | 0.119 | -1.574 | 0.115 | |
| **Nutrient cycling** | Intercept | 0.050 | 0.071 | 0.699 | 0.485 | 0.454 |
| | Duration (years) | 0.063 | 0.022 | 2.816 | 0.005 | |
| **Climate regulation** | Intercept | 0.032 | 0.120 | 0.267 | 0.790 | 0.008 |
| | Duration (years) | 0.015 | 0.022 | 0.694 | 0.488 | |

Extended Data Table 5. Result of the multinomial regression models predicting the probability of trade-off outcomes of agricultural diversification practice over time, with crop yield vs. other response variables, in overall and individually.

| Categories of response variable | Response level | Predictors | Log-Odds | std.error | statistic | p.value | RRR | nobs |
|---|---|---|---|---|---|---|---|---|
| Yield vs. all services | neg-pos | (Intercept) | 0.566 | 0.015 | 38.813 | 0.000 | 1.762 | 1448 |
| Yield vs. all services | neg-pos | Duration | 0.098 | 0.005 | 17.839 | 0.000 | 1.103 | 1448 |
| Yield vs. all services | pos-neg | (Intercept) | 0.647 | 0.016 | 39.246 | 0.000 | 1.909 | 1448 |
| Yield vs. all services | pos-neg | Duration | -0.174 | 0.008 | -20.521 | 0.000 | 0.841 | 1448 |
| Yield vs. all services | pos-pos | (Intercept) | 1.685 | 0.013 | 125.610 | 0.000 | 5.390 | 1448 |
| Yield vs. all services | pos-pos | Duration | 0.006 | 0.005 | 1.112 | 0.266 | 1.006 | 1448 |
| Yield vs. Financial profitability | neg-pos | (Intercept) | 0.827 | 0.049 | 16.976 | 0.000 | 2.287 | 98 |
| Yield vs. Financial profitability | neg-pos | Duration | -0.097 | 0.015 | -6.257 | 0.000 | 0.908 | 98 |
| Yield vs. Financial profitability | pos-neg | (Intercept) | 0.380 | 0.075 | 5.073 | 0.000 | 1.463 | 98 |
| Yield vs. Financial profitability | pos-neg | Duration | -0.228 | 0.038 | -6.051 | 0.000 | 0.796 | 98 |
| Yield vs. Financial profitability | pos-pos | (Intercept) | 1.688 | 0.061 | 27.767 | 0.000 | 5.407 | 98 |
| Yield vs. Financial profitability | pos-pos | Duration | -0.446 | 0.033 | -13.711 | 0.000 | 0.640 | 98 |
| Yield vs. Biodiversity | neg-pos | (Intercept) | -0.461 | 0.083 | -5.543 | 0.000 | 0.631 | 93 |
| Yield vs. Biodiversity | neg-pos | Duration | 0.890 | 0.095 | 9.357 | 0.000 | 2.436 | 93 |
| Yield vs. Biodiversity | pos-neg | (Intercept) | -1.687 | 0.148 | -11.411 | 0.000 | 0.185 | 93 |
| Yield vs. Biodiversity | pos-neg | Duration | 0.721 | 0.113 | 6.404 | 0.000 | 2.055 | 93 |
| Yield vs. Biodiversity | pos-pos | (Intercept) | -0.107 | 0.081 | -1.329 | 0.184 | 0.898 | 93 |
| Yield vs. Biodiversity | pos-pos | Duration | 0.902 | 0.095 | 9.501 | 0.000 | 2.465 | 93 |
| Yield vs. Pollination | neg-pos | (Intercept) | -1.887 | 0.371 | -5.093 | 0.000 | 0.151 | 11 |
| Yield vs. Pollination | neg-pos | Duration | 1.342 | 0.188 | 7.128 | 0.000 | 3.828 | 11 |
| Yield vs. Pollination | pos-neg | (Intercept) | -3.758 | 0.541 | -6.952 | 0.000 | 0.023 | 11 |
| Yield vs. Pollination | pos-neg | Duration | 1.597 | 0.226 | 7.066 | 0.000 | 4.936 | 11 |
| Yield vs. Pollination | pos-pos | (Intercept) | -101.12 | 135.575 | -0.746 | 0.456 | 0.000 | 11 |
| Yield vs. Pollination | pos-pos | Duration | 1.787 | 0.920 | 0.802 | 0.422 | 5.973 | 11 |
| Yield vs. Pest control | neg-pos | (Intercept) | 0.715 | 0.054 | 13.323 | 0.000 | 2.045 | 55 |
| Yield vs. Pest control | neg-pos | Duration | -0.106 | 0.026 | -4.028 | 0.000 | 0.899 | 55 |
| Yield vs. Pest control | pos-neg | (Intercept) | -1.127 | 0.119 | -9.492 | 0.000 | 0.324 | 55 |
| Yield vs. Pest control | pos-neg | Duration | 0.136 | 0.039 | 3.515 | 0.000 | 1.146 | 55 |
| Yield vs. Pest control | pos-pos | (Intercept) | 0.182 | 0.066 | 2.778 | 0.005 | 1.200 | 55 |
| Yield vs. Pest control | pos-pos | Duration | -0.098 | 0.034 | -2.908 | 0.004 | 0.906 | 55 |
| Yield vs. Water regulation | neg-pos | (Intercept) | 0.362 | 0.051 | 7.134 | 0.000 | 1.436 | 72 |
| Yield vs. Water regulation | neg-pos | Duration | -0.033 | 0.023 | -1.471 | 0.141 | 0.967 | 72 |
| Yield vs. Water regulation | pos-neg | (Intercept) | 0.300 | 0.052 | 5.739 | 0.000 | 1.349 | 72 |
| Yield vs. Water regulation | pos-neg | Duration | -0.037 | 0.024 | -1.569 | 0.117 | 0.964 | 72 |
| Yield vs. Water regulation | pos-pos | (Intercept) | 1.524 | 0.041 | 37.022 | 0.000 | 4.589 | 72 |
| Yield vs. Water regulation | pos-pos | Duration | -0.010 | 0.017 | -0.602 | 0.547 | 0.990 | 72 |
| Yield vs. Soil fertility | neg-pos | (Intercept) | 1.372 | 0.051 | 26.647 | 0.000 | 3.942 | 421 |
| Yield vs. Soil fertility | neg-pos | Duration | -0.004 | 0.011 | -0.388 | 0.698 | 0.996 | 421 |
| Yield vs. Soil fertility | pos-neg | (Intercept) | 1.889 | 0.053 | 35.354 | 0.000 | 6.611 | 421 |
| Yield vs. Soil fertility | pos-neg | Duration | -0.431 | 0.019 | -23.110 | 0.000 | 0.650 | 421 |
| Yield vs. Soil fertility | pos-pos | (Intercept) | 2.846 | 0.050 | 56.623 | 0.000 | 17.219 | 421 |
| Yield vs. Soil fertility | pos-pos | Duration | -0.133 | 0.011 | -12.535 | 0.000 | 0.876 | 421 |
| Yield vs. Nutrient cycling | neg-pos | (Intercept) | 1.557 | 0.057 | 27.435 | 0.000 | 4.746 | 357 |
| Yield vs. Nutrient cycling | neg-pos | Duration | -0.010 | 0.012 | -0.844 | 0.399 | 0.990 | 357 |
| Yield vs. Nutrient cycling | pos-neg | (Intercept) | 1.334 | 0.064 | 20.796 | 0.000 | 3.796 | 357 |
| Yield vs. Nutrient cycling | pos-neg | Duration | -0.373 | 0.028 | -13.495 | 0.000 | 0.689 | 357 |
| Yield vs. Nutrient cycling | pos-pos | (Intercept) | 2.878 | 0.056 | 51.574 | 0.000 | 17.773 | 357 |
| Yield vs. Nutrient cycling | pos-pos | Duration | -0.122 | 0.012 | -9.901 | 0.000 | 0.885 | 357 |
| Yield vs. Carbon sequestration | pos-neg | (Intercept) | 0.005 | 0.055 | 0.083 | 0.934 | 1.005 | 72 |
| Yield vs. Carbon sequestration | pos-neg | Duration | -0.511 | 0.045 | -11.256 | 0.000 | 0.600 | 72 |
| Yield vs. Carbon sequestration | pos-pos | (Intercept) | 0.834 | 0.019 | 42.850 | 0.000 | 2.302 | 72 |
| Yield vs. Carbon sequestration | pos-pos | Duration | -0.100 | 0.004 | -22.594 | 0.000 | 0.905 | 72 |
| Yield vs. Climate regulation | neg-pos | (Intercept) | 0.164 | 0.031 | 5.334 | 0.000 | 1.178 | 269 |
| Yield vs. Climate regulation | neg-pos | Duration | -0.115 | 0.020 | -5.806 | 0.000 | 0.891 | 269 |
| Yield vs. Climate regulation | pos-neg | (Intercept) | 0.322 | 0.026 | 12.419 | 0.000 | 1.379 | 269 |
| Yield vs. Climate regulation | pos-neg | Duration | -0.042 | 0.015 | -2.863 | 0.004 | 0.959 | 269 |
| Yield vs. Climate regulation | pos-pos | (Intercept) | 0.839 | 0.022 | 38.336 | 0.000 | 2.314 | 269 |
| Yield vs. Climate regulation | pos-pos | Duration | -0.012 | 0.012 | -1.033 | 0.302 | 0.988 | 269 |

**Supplementary materials**

**PRISMA (Preferred Reporting Items for Systematic Reviews and Meta-Analyses)**

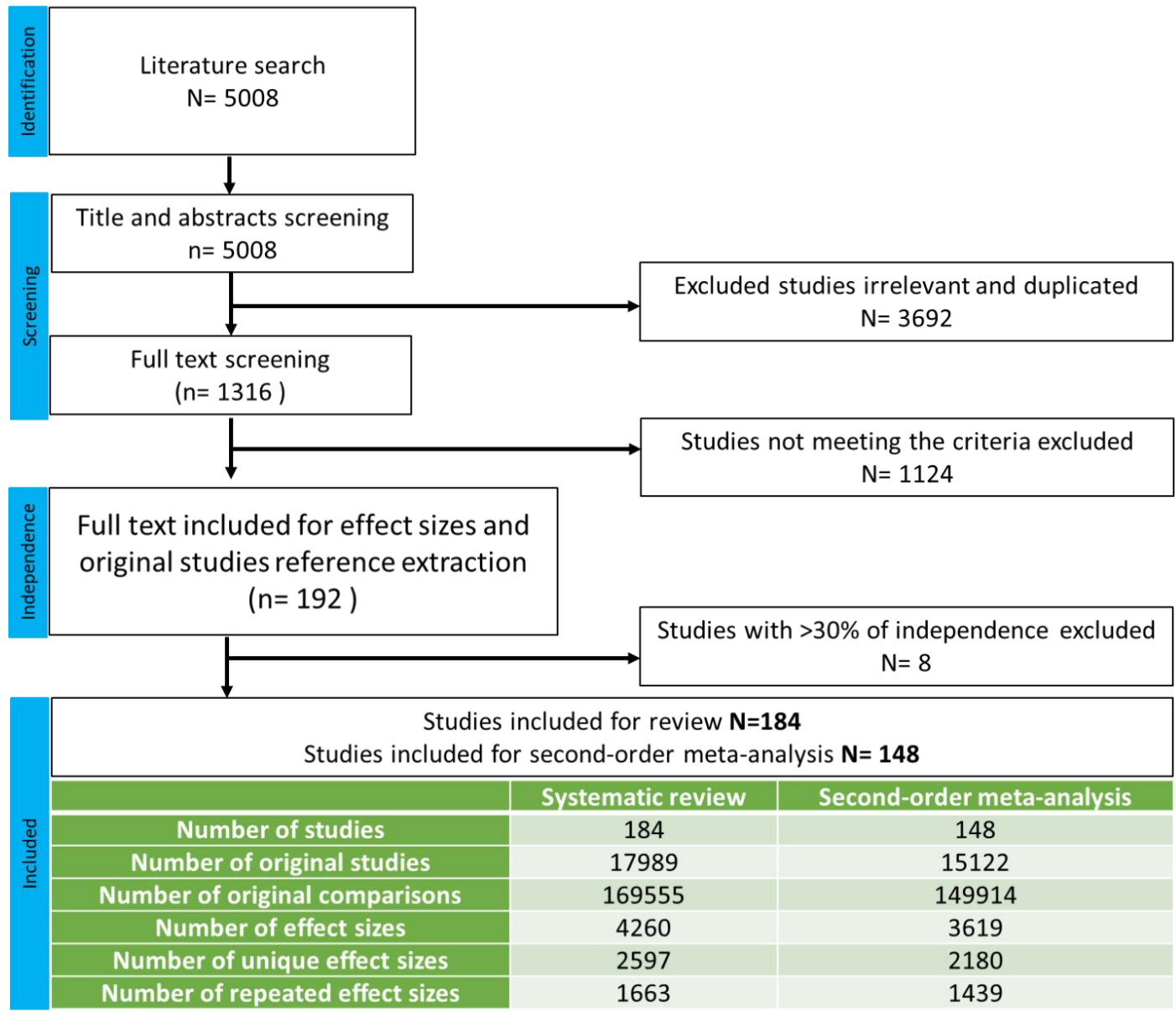

Supplementary Figure 1. PRISMA flow chart detailing the process of screening and extracting data from the studies obtained from the literature search.

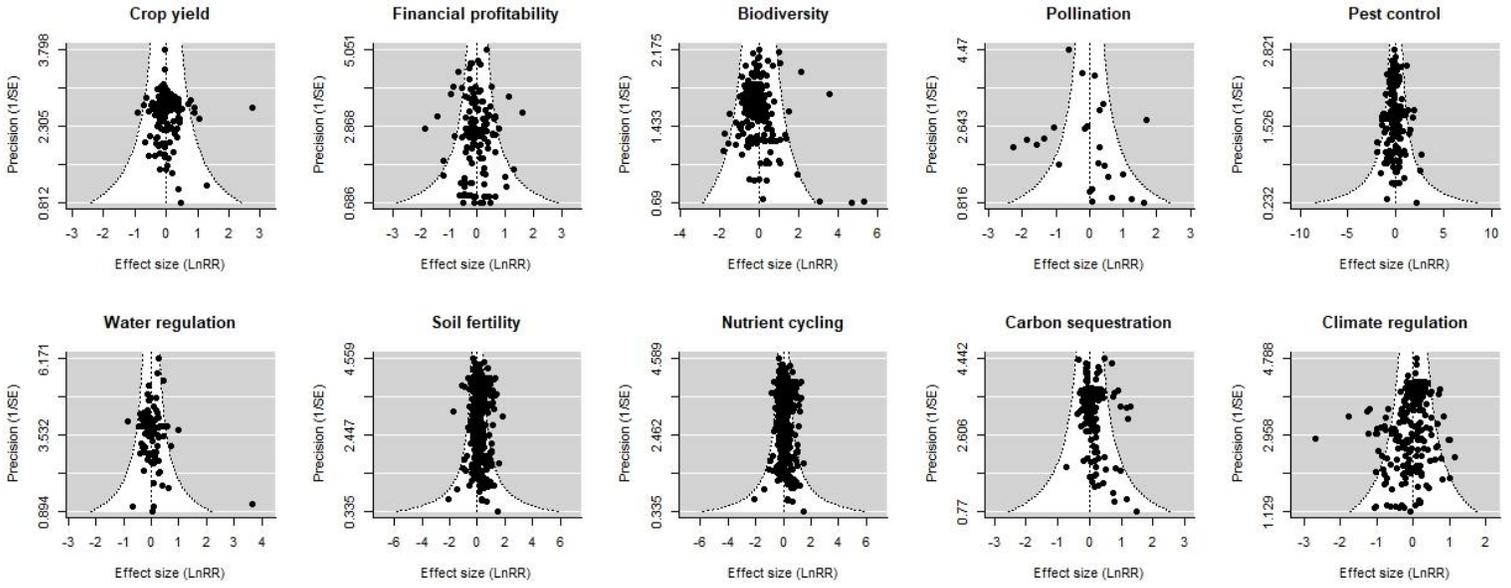

Supplementary Figure 2. Publication bias based on funnel plots: Effect size (lnRR) and the precision or inverse standard error (1/SE) for each category of the effect sizes.

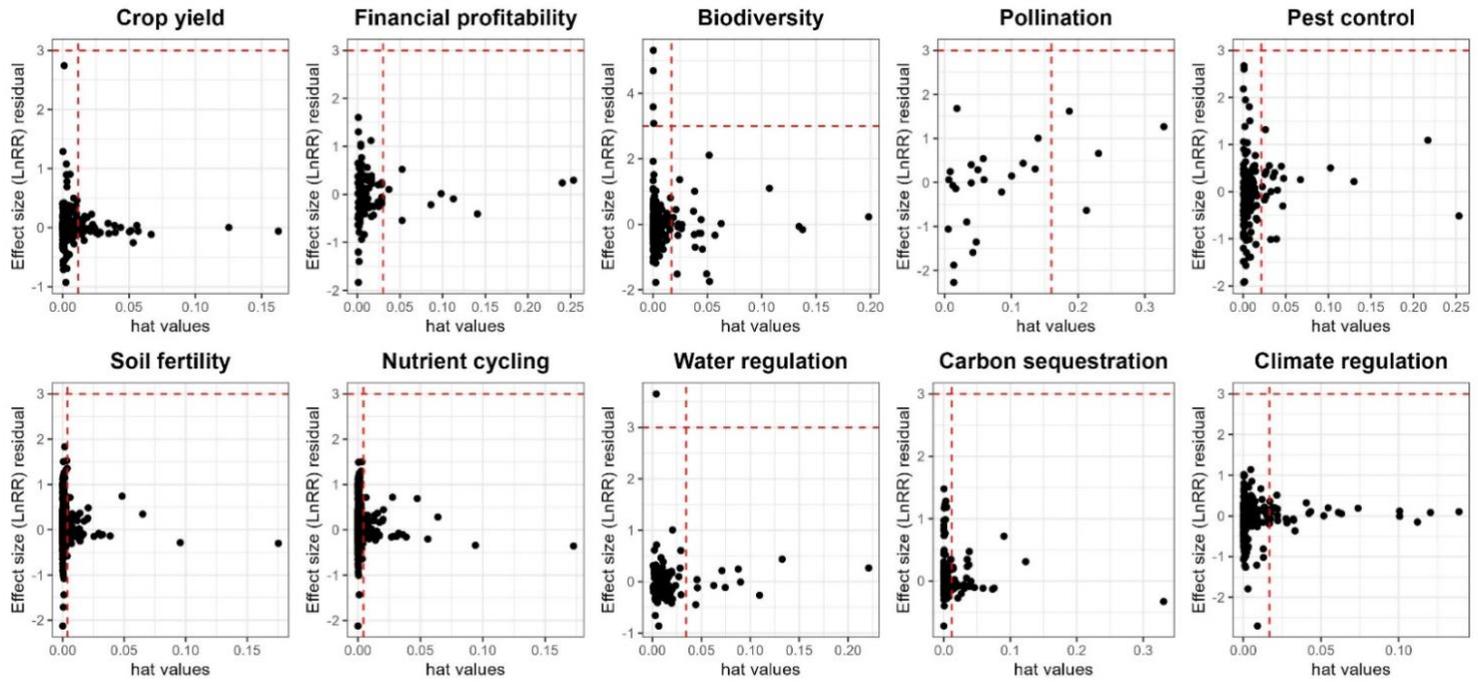

Supplementary Figure 3. Publication bias based on the influence of outliers: effect size r(LnRR) residual and the hat value for each category of the effect sizes.

**A) Overall diversificaiton**

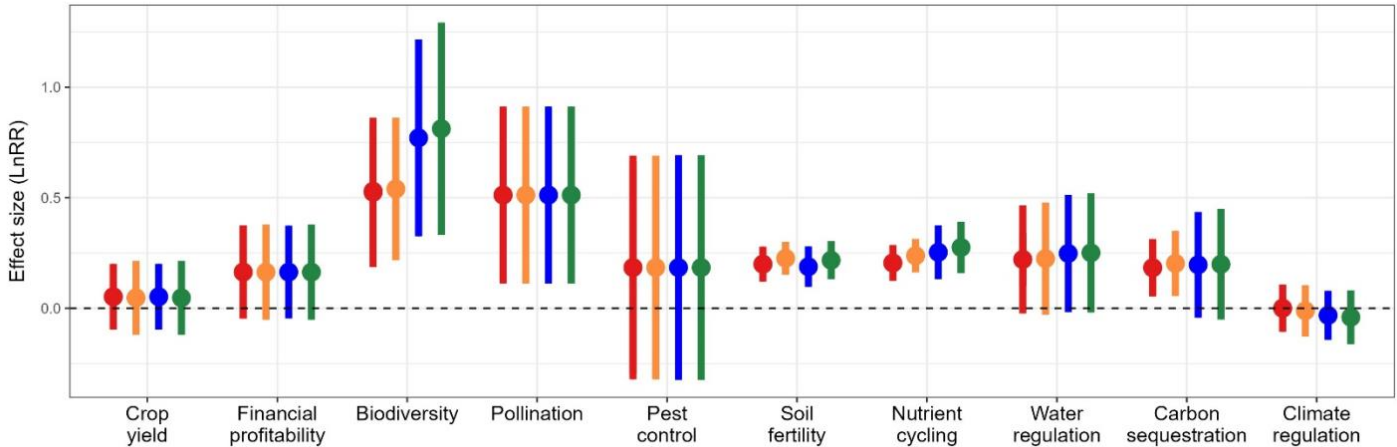

**B) Diversificaiton over time**

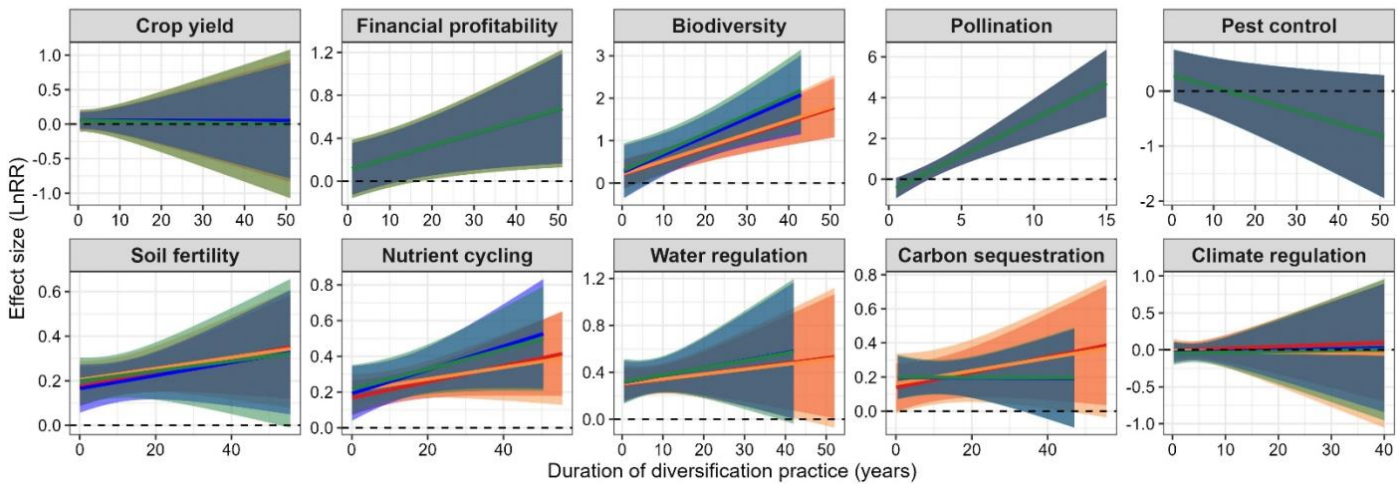

Supplementary Figure 4. Sensitivity analysis for the second order meta-analysis: comparison of model results from the effect of diversification over time using the i) full dataset to the model results, ii) full dataset without repeated responses, iii) dataset with high quality studies only and iv) dataset with high quality studies only and no repeated studies. Some variables only show the regression slopes for high quality studies and no repeated responses (green): financial profitability - no repeated and only few low-quality studies; pollination - no repeated responses and all high-quality studies; pest control - few repeated responses and all high quality.

Supplementary Table 1. Research strings to extract studies from Web of science and Scopus.

| | |
|---|---|
| Agricultural diversification systems | crop* OR "noncrop" OR intercropping OR "inter-cropping" OR "inter cropping" OR compost* OR till* OR "vegetation strip*" OR agroforest* OR inoculat* OR landscape OR organic OR fallow OR conventional OR fert* OR reduce* OR rotation OR "catch crop" OR amend* |
| Ecosystem services, yield, and financial profitability | divers* OR soil* OR biomass* OR water* OR pollutant* OR sediment* OR fodder OR emission* OR carbon* OR climat* OR pest* OR "biocontrol" OR weed* OR pollinat* OR fert* OR energ* OR resistance OR productiv* OR yield* OR "economic*" OR "profitability" OR "profit*" "econ*" OR "financ*" OR "gross returns" OR "gross income" OR "gross margin" OR "input*" OR "net return" OR "net income" |
| Type of studies | "meta-analysis" OR "metaanalysis" OR "meta analysis" |
| Publication date | Until 2022-12-31 |

Supplementary Table 2. Response variables and the categories to quantify the effects of diversification practices on socioeconomic, biological community, soil quality, and climate change mitigation factors adapted from Sánchez et al., 2022; Tamburini et al., 2020

| | Variables' categories | Included response variables |
|---|---|---|
| Socioeconomic factors | Crop yield | Crop yield; Coffee yield; Productivity; Root biomass; Shoot biomass; Biomass; Main yield; Belowground biomass; Aboveground productivity; Above ground biomass; Yield loss; Total biomass; Fruit productivity; Timber productivity; System yield |
| | Financial profitability | Product price; Gross margins; Equipment costs; Labor costs; Herbicide costs; Management costs; Net return; Total cost; Gross return; Benefit/cost ratio; Gross income; Production costs; System revenue; Net present value; Price premiums; Gross revenue; Net revenue |
| Biological community | Biodiversity | Microbes richness; Actinomyces abundance; Animal species richness; Arthropods evenness; Arthropods richness; Autotrophs diversity; Bacteria abundance; Bacterial community diversity; Bacterial richness; Bird abundance; Bird and mammal diversity; Bird diversity; Bird richness; Community structure; Decomposers diversity; Decomposers richness; Detritivore richness; Earthworm abundance; Earthworm density; Earthworm richness; Forest species richness; Frugivores abundance; Frugivores diversity; Frugivores richness; Functional diversity; Fungal community diversity; Fungi abundance; Fungi richness; Fungi taxonomic diversity; Granivores diversity; Insectivores diversity; Insects abundance; Insects richness; Microbial diversity; Mite density; Natural enemy diversity; Nematode density; Nematode richness; Omnivore richness; Omnivores diversity; Phylogenetic diversity; Plant abundance; Plant species |
| | | richness; Pollinator diversity; Pollinator richness; Predator richness; Prokaryotes taxonomic diversity; Protozoa richness; Richness AMF; Richness index; Shannon index; Species diversity; Species richness; Species richness and abundance; Springtail density |
| | Pollination | Pollinator richness; Pollinator diversity; Pollinator abundance; Pollinator evenness; Bee pollination |
| | Pest control | Arthropods pest abundance; Biocontrol; Disease damage; Herbivore abundance; Herbivore diversity; Herbivore evenness; Herbivore richness; Incidence of diseases; Incidence of pests; Insectivores diversity; Natural enemy diversity; Nematode damage; Omnivore richness; Omnivores diversity; Parasitoid abundance; Parasitoid richness; Pathogen control; Pathogen infestation; Pest damage; Pest infestation; Pests diversity; Plant damage; Predator abundance; Predator evenness; Predator richness; Proportion of healthy berries; Proportion of healthy leaves; Total natural enemy response; Total pest response; Weed abundance; Weed biomass; Weed control; Weed density; Weed infestation; Weed suppression; Weeds; Weeds diversity |
| Soil quality | Water regulation | Aggregate stability; Infiltration rate; N leaching; N leaching loss; N nitrate; N runoff; NO3 leaching; PAW - Plant Available Water; PWP - Permanent Wilting Point; Root length; Runoff; Sediment yield; Soil bulk density; Soil moisture; Soil porosity; Soil runoff; Soil saturated hydraulic conductivity; Soil water; Total P load; Total P to water; Volumetric water content; Water productivity; Water regulation; Water runoff; Water sediment; Water use; Water use efficiency |
| | Soil fertility | SOC; Abundance AMF; Actinomyces abundance; Aggregate stability; Alkali-hydrolysable nitrogen; Available K; Available N; Available P; Bacteria abundance; Bacteria biomass; Bacterial richness; Belowground carbon; Beta-glucosidase; Carbon sequestration; Catalase; Crop yield; Decomposers abundance; Decomposers diversity; Decomposers richness; Dehydrogenase; Detritivore abundance; Detritivore richness; Dissolved organic C; Earthworm abundance; Earthworm biomass; Earthworm density; Earthworm richness; Factor productivity of N; Fungi abundance; Fungi biomass; Fungi richness; Hydrolysable N; Invertase; Macroaggregates; Mean weight diameter of aggregates; Microaggregates; Microbes abundance; Microbes richness; Microbial biomass C; Microbial biomass N; Microbial diversity; Mite density; Mycorrhizal colonisation; N export; N leaching loss; N loss; N loss reduction; N nitrate; N use efficiency; Nematode abundance; Nematode density; Nematode richness; Nitrate N; Olsen P; Organic matter; P loss reduction; P use efficiency; PAW; Peroxidase; Phosphatase; Plant K uptake; Plant N uptake; Plant P uptake; Porosity; Potentially mineralizable N; PWP; Ratio MBC:MBN; Sediment yield; SHC; SOC stocks; Soil aggregation; Soil available K; Soil available P; Soil bulk density; Soil C; Soil C stocks; Soil C:N ratio; Soil C:P ratio; Soil catalase; Soil CEC; Soil EC; Soil enzyme activities; Soil fertility; Soil K; Soil loss; Soil moisture; Soil N; Soil N:P ratio; Soil N04; Soil organic matter; Soil P; Soil pH; Soil porosity; Soil runoff; Soil saturated hydraulic conductivity; Soil sucrase; Soil urease; Soil water-stable aggregation; Springtail density; Total C; Total C in system; Total K; Total N; Total N accumulation; Total P; Total SOC; Water sediment; Water stable aggregates |
| | Nutrient cycling | SOC; Actinomyces abundance; Alkali-hydrolysable nitrogen; Available K; Available N; Available P; Bacteria abundance; Bacteria biomass; Bacterial community diversity; Bacterial richness; Carbon sequestration; Colonization root length; Decomp abundance; Decomposers diversity; Decomposers richness; Detritivore abundance; Detritivore evenness; Detritivore richness; Dissolved organic C; Dissolved organic N; Earthworm density; Earthworm abundance; Earthworm richness; Fungal community diversity; Fungi abundance; Fungi biomass; Fungi richness; Hydrolysable N; Microbes abundance; Microbes richness; Microbial biomass; Microbial biomass C; Microbial biomass N; Microbial diversity; Microbial Shannon index; Mite density; Mycorrhizal colonisation; N content; N export; N leaching; N leaching loss; N loss reduction; N nitrate; N runoff; N use efficiency; Nematode abundance; Nematode density; Nematode richness; Nitrate N; Organic matter; P concentration; P content; P loss reduction; Plant P uptake; Potentially mineralizable N; Ratio MBC:MBN; Root N concentration; Root nodules; Root P concentration; SOC stocks; Soil available K; Soil available P; Soil C; Soil C stocks; Soil K; Soil N; Soil organic matter; Soil P; Springtail density; Total C; Total C in system; Total K; Total N; Total N accumulation; Total P; Total SOC |
| Climate regulation | Carbon sequestration | SOC; Above ground carbon; Carbon sequestration; SOC stocks; Belowground carbon; Total C; Total SOC; Soil C stocks; Soil C; Dissolved organic C; Soil organic matter; Total C in system |
| | Climate regulation | CH4 emission; N use efficiency; CO2 emission; NO2 emission; Plant N uptake; N2O emission; NH3 emission; Global warming potential; Biological N2 fixation; Factor productivity of N; Soil respiration; Total N accumulation; Soil NH5; N2 fixation; N loss |

Supplementary Table 3. Details of diversification practices under each category of agricultural diversification, similar to Tamburini et al., 2020

| Practices categories | Practice details |
|---|---|
| Non-crop diversification | Landscape complexity; Mixed non crop species; Biofuel feed-stock; Integrating trees; Hedgerow; Afforestation; Buffer strips; Grass cover; Embedded natural; Sod cultivation; Animal-crop symbiosis; Fallow practices; Living mulch; Livestock incorporation; Agroforestry |
| Organic amendment | Biosolids application; Residue retention; Manure application; Straw amendment; Organic amendment; Biochar amendment; Compost fertilizer; Organic fertilizer |

| Crop diversification | Crop rotation; Agroforestry; Cover cropping; Strip cropping; Intercropping; Cultivar mixture; Polyculture; Catch cropping; Mixed cropping |
|---|---|
| Reduced tillage | Reduced tillage; No tillage; Direct seeding; Subsoiling tillage; Minimum tillage |
| Organic farming | Organic farming; Organic management |
| Inoculation | AMF inoculation; EMF inoculation; Rhizobia co-inoculation |

Supplementary Table 4. List of information for assessing the quality of each meta-analysis study included in this study. Each information was scored one when absent and two when present. The score of each study was equal to the total scores of all information.

| | Quality assessment criteria | Score |
|---|---|---|
| Methods | Literature search method provided | No=1 , Yes=2 |
| | Inclusion/exclusion criteria of original studies reported | Total score=16 |
| | Definition of the control group included | Quality Assessment |
| | Heterogeneity assessment included | 8-13 >> Low quality |
| | Weighting procedure reported | 14-16 >> High quality |
| Results | Number of original studies reported | |
| | Effect size average and Cis included | |
| | Sensitivity analysis control included | |

Supplementary Table 5. Results of Egger's regression test to detect asymmetry of funnel plots for each category of the response variables.

| | | estimate | se | zval | pval | ci.lb | ci.ub | Sign. |
|---|---|---|---|---|---|---|---|---|
| Crop yield | intrcpt | 0.043 | 0.0812 | 0.5297 | 0.5963 | -0.1162 | 0.2022 | |
| | mods | 0.2658 | 0.6918 | 0.3842 | 0.7008 | -1.0901 | 1.6217 | |
| Finance profitability | intrcpt | 0.1157 | 0.1629 | 0.7103 | 0.4775 | -0.2035 | 0.4349 | |
| | mods | 0.157 | 0.3293 | 0.4769 | 0.6335 | -0.4883 | 0.8023 | |
| Biodiversity | intrcpt | 0.5037 | 0.2122 | 2.3742 | 0.0176 | 0.0879 | 0.9196 | * |
| | mods | 0.15 | 0.6157 | 0.2436 | 0.8075 | -1.0568 | 1.3568 | |
| Pollination | intrcpt | -0.1903 | 0.246 | -0.7736 | 0.4392 | -0.6724 | 0.2918 | |
| | mods | 1.5605 | 0.5931 | 2.6312 | 0.0085 | 0.3981 | 2.723 | ** |
| Pest control | intrcpt | 0.2979 | 0.2731 | 1.0905 | 0.2755 | -0.2375 | 0.8332 | |
| | mods | -0.3606 | 0.4025 | -0.8958 | 0.3703 | -1.1495 | 0.4283 | |
| Water regulation | intrcpt | 0.1264 | 0.075 | 1.6858 | 0.0918 | -0.0205 | 0.2733 | . |
| | mods | 1.0992 | 0.43 | 2.5562 | 0.0106 | 0.2564 | 1.942 | * |
| Soil fertility | intrcpt | 0.1647 | 0.0404 | 4.0742 | <.0001 | 0.0855 | 0.2439 | *** |
|  | mods | 0.541 | 0.2586 | 2.0924 | 0.0364 | 0.0342 | 1.0479 | * |
| Nutrient cycling | intrcpt | 0.175 | 0.0455 | 3.8479 | 0.0001 | 0.0859 | 0.2642 | *** |
|  | mods | 0.4493 | 0.2736 | 1.6424 | 0.1005 | -0.0869 | 0.9855 |  |
| Carbon sequestration | intrcpt | 0.1422 | 0.0728 | 1.9529 | 0.0508 | -0.0005 | 0.2848 | . |
|  | mods | 0.8835 | 0.4404 | 2.0061 | 0.0448 | 0.0203 | 1.7467 | * |
| Climate regulation | intrcpt | 0.0594 | 0.0674 | 0.8808 | 0.3784 | -0.0728 | 0.1915 |  |
|  | mods | -0.7442 | 0.459 | -1.6216 | 0.1049 | -1.6438 | 0.1553 |  |

Supplementary Table 6. Results of fail-safe N analysis based on Rosenthal method for each category of the response variables.

|  | Fail-safe N | Observed significant level |
|---|---|---|
| Crop yield | 382509 | <.0001 |
| Financial profitability | 1259 | <.0001 |
| Biodiversity | 713221 | <.0001 |
| Pollination | 348 | <.0001 |
| Water regulation | 234372 | <.0001 |
| Pest control | 1483 | <.0001 |
| Soil fertility | 15021430 | <.0001 |
| Nutrient cycling | 11475248 | <.0001 |
| Carbon sequestration | 2709693 | <.0001 |
| Climate regulation | 8611 | <.0001 |

Supplementary Table 7. Results of publication bias test based on Likelihood Ratio Test for each category of the response variables.

|  | $X^2$ | p-value |
|---|---|---|
| Crop yield | 0.1652429 | 0.68437 |
| Financial profitability | 1.992073 | 0.15812 |
| Biodiversity | 2.227789 | 0.39767 |
| Pollination | 0.1792251 | 0.67204 |
| Pest control | 0.7644067 | 0.38195 |
| Water regulation | 0.07537393 | 0.78367 |
| Soil fertility | 1.232491 | 0.26692 |
| Nutrient cycling | 3.145582 | 0.076133 |
| Carbon sequestration | 2.71357 | 0.099498 |
| Climate regulation | 2.775196 | 0.28873 |